\documentclass[twocolumn,twocolappendix]{aastex63}%%% For overleaf, two-column

\graphicspath{{./}{./F/}{./F/ps/}}

\usepackage[]{txfonts}
\usepackage[T1]{fontenc}

\newcommand{\eqref}[1]{Equation (\ref{#1})}
\newcommand{\dfrac}[2]{ {\displaystyle\frac{#1}{#2}} }
\newcommand{\pr}[1]{\ensuremath{\left(#1\right)} }

\newcommand{\pf}[2]{\ensuremath{\left( \dfrac{#1}{#2} \right)} }

\newcommand{\etaA}{\eta_{\rm A} }

\newcommand{\Am}{{\rm Am} }

\newcommand{\z}{$z$}
\renewcommand{\d}{{\rm d}}

\newcommand{\dz}{\,{\rm d}z}

\newcommand{\alppz}{\overline{\alpha_{z \phi}}}
\newcommand{\SigmaAm}{\Sigma_{\rm Am=0.3}}
\newcommand{\Sigmaheat}{\Sigma_{\rm heat}}
\newcommand{\Sigmamid}{\Sigma_{\rm mid}}

\newcommand{\K}{\ensuremath{{\rm \,K }} }
\newcommand{\au}{\ensuremath{{\rm \,au }}}
\newcommand{\yr}{\ensuremath{\rm \,yr }}
\newcommand{\Msun}{\ensuremath{\rm \,M_{\sun}}}
\newcommand{\Lsun}{\ensuremath{\rm \,L_{\sun}}}
\newcommand{\Mpyr}{ \ensuremath{\rm \,\Msun \,\yr^{-1} }}
\newcommand{\cmcmg}{\ensuremath{\rm \,cm^{2} \, g^{-1} }}
\newcommand{\gcmcm}{\ensuremath{\rm \,g \, cm^{-2} }}

\newcommand{\tcrossMHD}{0.6 }
\newcommand{\tcrossvisc}{10 }

\definecolor{darkorange}{rgb}{1.0, 0.55, 0.0}

\defcitealias{Mori2019Temperature-Str}{MBO19}

\begin{document}
\title{
Evolution of the Water Snow Line in Magnetically Accreting Protoplanetary Disks 
}

\shortauthors{Mori et al.}

%%%%%%%%%%%%%%%%%%%%%%%%
%% Affiliation 
\def\myemail{mori.s@astr.tohoku.ac.jp}
\def\utokyo{Department of Astronomy, The University of Tokyo, 7-3-1 Hongo, Bunkyo-ku, Tokyo 113-0033, Japan}
\def\tohoku{Astronomical Institute, Tohoku University, 6-3 Aramaki, Aoba-ku, Sendai 980-8578, Japan}
\def\kurume{Department of Physics, Kurume University, 67 Asahi-machi, Kurume, Fukuoka 830-0011, Japan}
\def\titech{Department of Earth and Planetary Sciences, Tokyo Institute of Technology, Meguro-ku, Tokyo 152-8551, Japan}
\def\tsinghua{Institute for Advanced Study and Department of Astronomy, Tsinghua University, Beijing 100084, China}

\author[0000-0002-7002-939X]{Shoji Mori}       \affiliation{\utokyo}   \affiliation{\tohoku; \href{mailto:\myemail}{\rm \myemail}}
\author[0000-0002-1886-0880]{Satoshi Okuzumi}  \affiliation{\titech}
\author[0000-0002-1932-3358]{Masanobu Kunitomo}  \affiliation{\kurume}
\author[0000-0001-6906-9549]{Xue-Ning Bai}     \affiliation{\tsinghua}  

%%%%%%%%%%%%%%%%%%%%%%%%%%%%

%%%%%%%%%%%%%%%%%%%%%%%%%%%%%%%%%%%%%%%%%%%%
\begin{abstract}
The low water content of the terrestrial planets in the solar system suggests that the protoplanets formed within the water snow line.
Accurate prediction of the snow line location moving with time provides a clue to constrain the formation process of the planets.
In this paper, we investigate the migration of the snow line in protoplanetary disks whose accretion is controlled by laminar magnetic fields,
which have been proposed by various nonideal magnetohydrodynamic (MHD) simulations.
We propose an empirical model of the disk temperature based on our nonideal MHD simulations, which show that the accretion heating is significantly less efficient than in turbulent disks,
 and calculate the snow line location over time. 
We find that the snow line in the magnetically accreting laminar disks moves inside the current Earth's orbit within 1 Myr after star formation,
whereas the time for the conventional turbulent disk is much longer than 1 Myr.
This result suggests that either the rocky protoplanets formed in such an early phase of the disk evolution, 
or the protoplanets moved outward to the current orbits after they formed close to the protosun.  
\end{abstract}% 
%%%%%%%%%%%%%%%%%%%%%%%%%%%%%%%%%%%%%%%%%%%%

%%%%%%%%%%%%%%%%%%%%%%%%%%%%%%%%%%%%%%%%%%%%%%%%%%%%%%%%%%%%%%%%%%%%%%%%%%%%%%

%%%%%%%%%%%%%%%%%%%%%%%%%%%%%%%%%%%%%%%%%%%%%%%
%%%%%%%%%%%%%%%%%%%%%%%%%%%%%%%%%%%%%%%%%%%%%%%
%%%%%%%%%%%%%%%%%%%%%%%%%%%%%%%%%%%%%%%%%%%%%%%
\section{Introduction}\label{sec:intro}

The terrestrial planets in the solar system are significantly drier than solids in the outer solar system.
The mass of the present Earth's ocean is only 0.02 \% of the Earth's mass \citep[e.g.,][]{charette2010volume}. 
Even considering the amount of water that may have been taken into the Earth's interior, 
the Earth's water content is at most 2 wt\% \citep{Nomura2014Low-Core-Mantle,Fei2017A-nearly-water-}.
In addition, current Venus has a dry atmosphere with a water content of $\sim$30 ppm \citep[e.g.,][]{Basilevsky2003The-surface-of-}, 
and its interior would only contain at most 10 \% of the Earth's water content \citep{Elkins-Tanton2007Volcanism-and-v}.
Mars is likely to contain only up to 10$^{-4}$ Mars' mass of water \citep{Kurokawa2014Evolution-of-wa}.
For Mercury, the only evidence of water is ice at the north pole, which is only 10$^{-8}$ Mercury's mass \citep{Lawrence2013Evidence-for-Wa}.
Therefore, their present water contents would be $\lesssim 1 $ wt\%.
In contrast, bodies originating in the outer solar system, such as comets and Neptune, have a higher water content over 10 wt\% \citep[e.g.,][]{Guillot2005THE-INTERIORS-O,AHearn2011EPOXI-at-Comet-,Rotundi2015Dust-measuremen}.

The water content at the time of the formation of the terrestrial planets may have been higher than the present value, but would still have been $\lesssim 1 $ wt\%.
For the Earth, $\sim 0.1 M_{\oplus}$ of water is hardly removed by stellar irradiation \citep{Machida2010Terrestrial-Pla,Hamano2013Emergence-of-tw} or giant impact \citep{Genda2005Enhanced-atmosp,Schlichting2015Atmospheric-mas,Biersteker2021Losing-Oceans:-}.
Therefore, the initial Earth would not have contained much more water than it does today.
The high hydrogen isotope ratio of present Venus suggests that massive dehydration has occurred in the past, 
but even so, the water content of early Venus is estimated to be about the same as the Earth \citep{Donahue1982Venus-Was-Wet:-}.
Early Mars may have retained up to 0.05 \% of the water in the Mars' mass \citep{Kurokawa2014Evolution-of-wa}.

The water content of a planet is largely determined by the location of the water snow line during planet formation.
The water snow line is defined as a radius in protoplanetary disks (PPDs) where water sublimates and condenses.
In typical PPDs, the water snow line lies where the gas temperature is $\sim$ 160--170 K \citep[e.g.,][]{Hayashi1981Structure-of-th}.
Dust inside the snow line is essentially dry, whereas dust outside the snow line can contain up to  $\sim$ 50 wt\% of water ice \citep{Lodders2003Solar-System-Ab}.
Planetesimals forming from icy dust can lose water to some extent through radiogenic heating by $^{26}$Al, 
but the probability that the dehydrating planetesimals form rocky planets with a significantly low water content ($<$1 wt\%) is still low with the initial solar abundance of $^{26}$Al \citep{Lichtenberg2019A-water-budget-}.
Even if rocky planets form outside the snow line from dehydrated planetesimals, they may acquire a significant amount of water afterward by capturing icy particles in the gas disk \citep{Sato2016On-the-water-de}. 
Therefore, it is more natural to consider that the terrestrial planets in the solar system formed inside the snow line and stayed there until the icy dust outside the snow line was cleared out.

The position of the snow line is determined by the disk's thermal structure.
In the simplest case where the disk is optically thin and its internal temperature is determined by direct stellar irradiation, the snow line lies at $\approx 3$ au for the present-day solar luminosity \citep{Hayashi1981Structure-of-th}.
However, such a simple model does not apply to young, optically thick PPDs that contain abundant dust grains.
Optically thick disks can receive stellar radiation only on their surface, and for this reason their interior temperature tends to be lower than that of optically thin disks \citep{Kusaka1970Growth-of-Solid,Chiang1997Spectral-Energy}.
In an optically thick disk that is passively irradiated by a Sun-like star, the snow line can lie within 1 au \citep{Sasselov2000On-the-Snow-Lin}.  

However, stellar irradiation is not the only heating mechanism for optically thick PPDs. They can also be heated internally when the gas accretes toward the central star and liberates its gravitational energy  \citep{Shakura1973Black-holes-in-,Lynden-Bell1974The-evolution-o}.
Accretion heating is most efficient when the energy is liberated deep inside the disk from which it takes a long time for the heat to escape.
This effect is sometimes called the blanketing effect because an optically thick material acts as a thermal blanket \citep[e.g.,][]{Milne1921Radiative-equil,Chandrasekhar1935The-radiative-e}. 

Previous studies for the snow line in optically thick PPDs \citep[e.g.,][]{Sasselov2000On-the-Snow-Lin,Garaud2007The-Effect-of-I,Oka2011Evolution-of-Sn,Zhang2015The-Evolution-o,Xiao2017Time-evolution-} commonly adopt the classical viscous accretion disk model that assumes vertically uniform viscosity.
In this model, the blanketing effect is particularly effective, pushing the snow line out to a few au from the central star for typical values of the disk accretion rate \citep{Sasselov2000On-the-Snow-Lin,Garaud2007The-Effect-of-I,Oka2011Evolution-of-Sn}.
This indicates that accounting for accretion heating is essential to infer where the snow line was when the embryos of the terrestrial planets formed in the solar nebula.

An important consequence of the vertically uniform kinematic viscosity assumed in the classical viscous model is 
that the accretion energy is locally dissipated, and hence is mainly deposited near the midplane, thus maximizing the thermal blanket effect.
The question is then whether such a vertical heating profile is expected in a realistic model of disk accretion. 
In fact, magnetohydrodynamic (MHD) accretion models for PPDs predict that energy dissipation mainly takes place {\it near the disk surface} rather than near the midplane.
Turbulence generated by magnetorotational instability \citep[MRI;][]{Balbus1991A-powerful-loca} has been thought to be the dominant energy dissipation mechanism. 
The dissipation profile is affected by nonideal MHD effects (Ohmic diffusion, Hall effect, and ambipolar diffusion) brought about by the weakly ionized nature of PPDs.
Strong Ohmic diffusion in the inner disk region suppresses the MRI turbulence around the midplane \citep{Gammie1996Layered-Accreti,Fleming2003Local-Magnetohy}, leading to a {\it dead zone}.
In that case, the energy dissipates in the surface MRI-active layers above the dead zone \citep{Hirose2011Heating-and-Coo}, rather than around the midplane.
Furthermore, ambipolar diffusion operates in the lower-density region and hence quenches even the turbulence in the upper layer \citep{Bai2013aWind-driven-Acc,Bai2013bWind-driven-Acc,Gressel2015Global-Simulati,Simon2015Magnetically-dr}, where the disk accretion is driven by the angular momentum removal by magnetic disk winds \citep[i.e., wind-driven accretion;][]{Bai2013aWind-driven-Acc}.

With the MRI being suppressed, heating is due to Joule dissipation rather than turbulent dissipation,
which is not co-spatial with the location where the accretion energy is liberated.
Moreover, a substantial fraction of the liberated energy is consumed for driving disk winds rather than leading to dissipation.
\citet{Mori2019Temperature-Str} \citepalias[hereafter ][]{Mori2019Temperature-Str} 
performed MHD simulations with all the three nonideal MHD effects producing laminar magnetized accreting disks,
and showed that the Joule heating occurs at high altitude and hence the accretion heating is much less efficient than in the viscous disk model\footnote{\citet{Bethune2020Electric-heatin} also investigated the vertical distribution of Joule heating with the nonideal MHD effects, but led to a different conclusion from \citetalias{Mori2019Temperature-Str}. 
In Section \ref{ssec:BL20}, based on the results of this paper, we discuss the differences between the results of \citetalias{Mori2019Temperature-Str} and \citet{Bethune2020Electric-heatin}.}.
It was also shown that the Hall effect amplifies and suppresses Joule heating when the magnetic field threading the disk is aligned and anti-aligned with the disk rotation axis, respectively \citepalias{Mori2019Temperature-Str}. 

In this paper, we show the temporal evolution of the location of the water snow line in magnetically accreting PPDs.
To do so, we construct a global model of disk temperature evolution in the terrestrial-planet-forming region in the disks, based on the results of \citetalias{Mori2019Temperature-Str}.
\citetalias{Mori2019Temperature-Str} suggested that even in the early phase of PPD evolution, 
the disk temperature of the inner regions will be determined by the stellar irradiation.
We explore this scenario in much greater detail.
In addition, we also discuss the impact of the migration of the snow line on the terrestrial planet formation.
The water snow line migrates with the temporal evolution of the disk's thermal structure \citep[e.g.,][]{Oka2011Evolution-of-Sn}. 
Thus, considering that protoplanets of the terrestrial planets should have formed within the snow line, the track of the snow line constrains their formation time and location.
We suggest that the formation time of rocky protoplanets is strongly constrained, 
or that the protoplanets have moved outward from the vicinity of the protosun to the present orbits after the formation.

This paper is organized as follows. 
In Section \ref{sec:methods}, we propose a new disk temperature model based on MHD simulation results.
In Section \ref{sec:SL}, we show the time evolution of the water snow line around a 1 $M_{\odot}$ star.
In Section \ref{sec:discuss}, we discuss the implications of our results for rock planet formation and other possible heating mechanisms neglected in this paper.

%%%%%%%%%%%%%%%%%%%%%%%%%%%%%%%%%%%%%%%%%%%%%%%
%%%%%%%%%%%%%%%%%%%%%%%%%%%%%%%%%%%%%%%%%%%%%%%
%%%%%%%%%%%%%%%%%%%%%%%%%%%%%%%%%%%%%%%%%%%%%%%
\section{MHD Wind-Driven Accretion Disk Model}\label{sec:methods}

\begin{figure*}
	\centering
	\includegraphics[width = 0.75\hsize,clip]{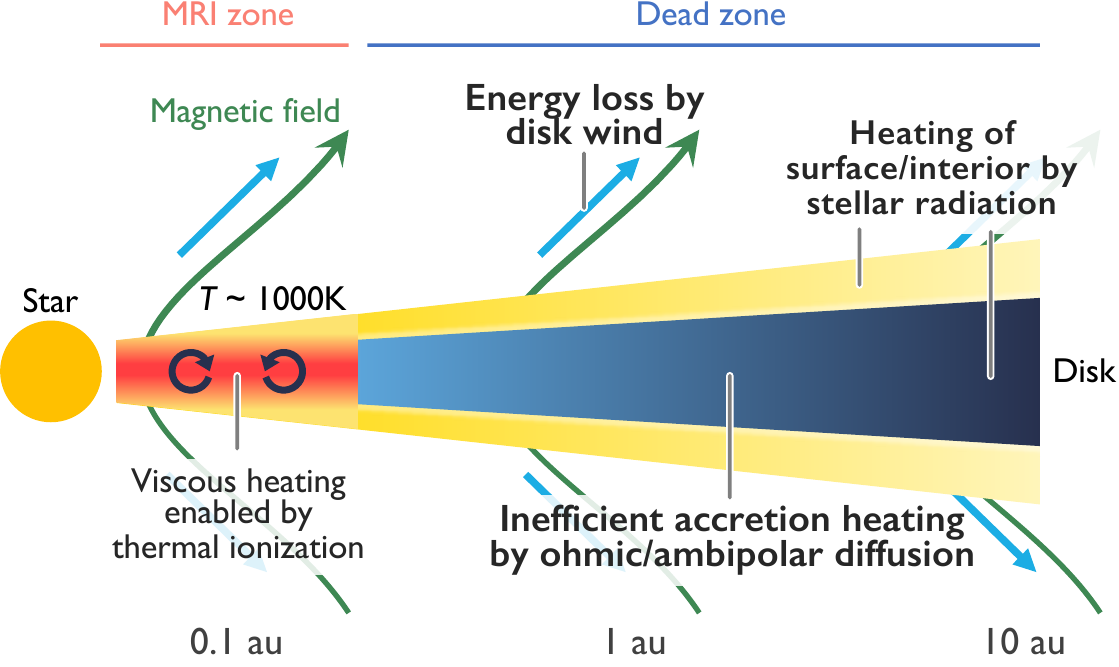}
	\caption{
	Schematic illustration of the expected thermal structure of PPDs.
	}
	\label{fig:tmap}
\end{figure*}

We divide the inner region of PPDs into two zones in terms of the thermal and turbulent states as summarized in Figure \ref{fig:tmap}: 
MRI zones and dead zones. 
The MRI zone is the very inner hot region \citep[$T\gtrsim 1000$ K;][]{Desch2015High-temperatur} where MRI-turbulence is induced by the thermal ionization and heats the gas with the turbulent viscosity.
Outside of the MRI zone is the dead zone, where the MRI is suppressed without thermal ionization, and cooler because of the inefficient accretion heating.  
In the following, we focus on the dead zone because we are interested in the evolution of the snow line where $T \sim 160~\rm K$.

%%%%%%%%%%%%%%%%%%%%%%%%%%%%%%%%%%%%%%%%%%%%%%%
\subsection{Density Structure and Accretion Rate}
\label{sec:Sigma}
%%%%%%%%%%%%%%%%%%%%%%%%%%%%%%%%%%%%%%%%%%%%%%%
We assume that the disk is vertically nearly isothermal and give the vertical gas density profile as 
\begin{equation}
    \rho(z) = \frac{\Sigma}{\sqrt{2\pi} H} \exp\left(-\frac{z^2}{2H^2}\right),
    \label{eq:rho}
\end{equation}
where $z$ is the distance from the midplane, 
$H$ is the scale height,
and $\Sigma$ is the gas surface density.
The scale height is related to the isothermal sound speed $c_{\rm s}$ and local Keplerian frequency $\Omega$
as $H = c_{\rm s}/\Omega$. 
The gas can indeed be regarded as vertically isothermal at low altitude below heat sources \citepalias[see][]{Mori2019Temperature-Str}. 

Accretion in the dead zone is assumed to be entirely driven by magnetic winds. In this limit, the mass accretion rate $\dot{M}$ can be written as 
\citep{Wardle2007Magnetic-fields,Fromang2013Local-outflows-}
\begin{equation}
    \dot{M} = \frac{4\pi r}{\Omega} w_{\rm wind},
    \label{eq:Mdot}
\end{equation}
where $r$ is the radial distance, and $w_{\rm wind}$ is the $z \phi$ component of the Maxwell stress exerted on both sides of the disk, which we call the wind stress.

In principle, $w_{\rm wind}$ depends on the net flux of large-scale magnetic fields threading the disk \citep{Bai2013aWind-driven-Acc,Bai2013Local-Study-of-,Bai2014Hall-effect-Con}, which is, however, highly uncertain. 
For this reason, we opt for parameterizing $w_{\rm wind}$ as \citep{Suzuki2016Evolution-of-pr}
\begin{equation}
    w_{\rm wind} = P_{\rm mid} \overline{\alpha_{z \phi}},
    \label{eq:w_wind}
\end{equation}
where $P_{\rm mid}$ is the gas pressure at the midplane and $\overline{\alpha_{z \phi}}$ 
is a dimensionless parameter that characterizes the level of the wind stress.
According to MHD simulations \citep{Bai2017Global-Simulati,Mori2019Temperature-Str}, typical values of $\overline{ \alpha_{z \phi}}$ range between $10^{-4}$--$10^{-2}$.
We take $\overline{ \alpha_{z \phi}} = 10^{-3}$ as a default value and vary $\overline{ \alpha_{z \phi}}$ in Section \ref{sssec:SL-Maxwell}.

The parameter $\overline{ \alpha_{z \phi}}$ should not be confused with the $\alpha$ parameter in the standard viscous disk model \citep{Shakura1973Black-holes-in-}. 
The former measures the stress vertically transporting angular momentum to the magnetic wind, 
whereas the latter measures the radial angular momentum transport due to the magnetic fields and turbulence within the disk. 
Moreover, in general, magnetized wind is much more efficient in driving disk accretion.
This can be found by noting that $\Sigma$ in steady state is related to $\dot{M}$ as  $\Sigma \sim \dot{M}/(\overline{ \alpha_{z \phi}} c_{\rm s} r)$ for wind-driven accretion (see \eqref{eq:Sigma} below) and $\Sigma \sim \dot{M}
/(\alpha c_s H)$ for viscous accretion.
Comparing the two expressions, the two disk models give the same value of $\Sigma$ for given $\dot{M}$ when $\alpha \sim (r/H) \overline{ \alpha_{z \phi}}$ \citep{Wardle2007Magnetic-fields,Bai2009Heat-and-Dust-i,Fromang2013MRI-driven-angu,Bai2017Global-Simulati}.

Assuming wind-driven accretion with $\overline{\alpha_{z \phi}}$ being constant, one can relate $\Sigma$ to $\dot{M}$. 
When the disk is isothermal except at high altitude, the midplane pressure can be written as 
$P_{\rm mid} = \Sigma c_{\rm s}^2/(\sqrt{2\pi} H) = \Sigma \Omega c_{\rm s}/\sqrt{2\pi}$.
Using this and Equations (\ref{eq:Mdot}) and (\ref{eq:w_wind}), we have 
\begin{equation}
	\Sigma = \frac{\dot{M}}{2 \sqrt{2 \pi} \overline{ \alpha_{z \phi}}  c_{\rm s} r}.
	\label{eq:Sigma}
\end{equation}

\begin{figure}
	\centering
	\includegraphics[width = \hsize,clip]{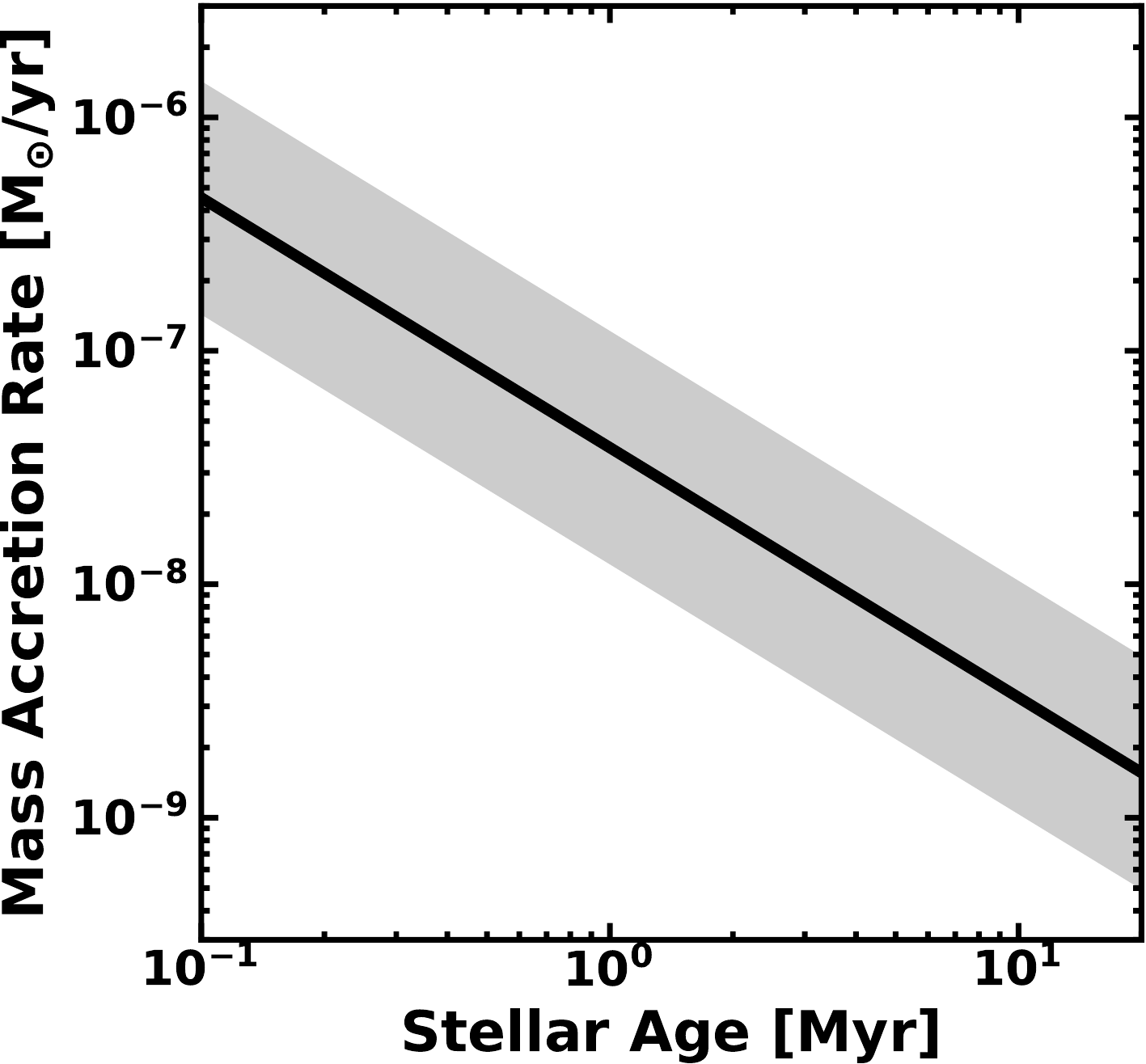}
	\caption{
	Empirical relation between the mass accretion rate $\dot{M}$ and stellar age $t$ (\eqref{eq:Mdot-t}; \citealt{Hartmann2016Accretion-onto-}).
	The shade shows the observational scatter of 0.5 dex.
	}
	\label{fig:Mdot}
\end{figure}

We assume that the disk is in a quasi-steady state, 
which follows from the assumption that $\dot{M}$ varies on timescales much longer than the local gas accretion timescale.
%the assumption that $\dot{M}$ varies on a much longer timescale than that of the radial gas flow at the place.
We also assume that mass accretion is dominant over mass loss. 
Thereby, the mass accretion rate $\dot{M}$ is radially constant \citep[see][]{Suzuki2016Evolution-of-pr,Bai2016Towards-a-Globa} and is equal to the gas accretion rate onto the central star.
We determine $\dot{M}$ as a function of stellar age $t$ 
using the empirical relation for the stellar mass $M = 1\,\rm M_{\sun}$ by \citet{Hartmann2016Accretion-onto-}
\footnote{We have rewritten the empirical relation of $\dot{M}$--$t$ for 0.7 $M_{\sun}$ stars (Equation (12) in \citet{Hartmann2016Accretion-onto-}) into that for 1.0  $M_{\sun}$ stars by using the empirical relation of $\dot{M}$--$M$ (Equation (11) in their paper).},  
\begin{equation}
	\dot{M} = 4\times10^{-8 \pm 0.5}  \pf{t}{1~{\rm Myr}}^{-1.07} ~{\rm M_{\sun}~{\rm yr}^{-1} } ,
	\label{eq:Mdot-t}
\end{equation}
which is shown in Figure \ref{fig:Mdot}.
Here, the stellar age $t$ is defined as the time after star formation is completed, 
i.e., after the end of the protostellar accretion phase and the arrival at the stellar birthline.

%%%%%%%%%%%%%%%%%%%%%%%%%%%%%%%%%%%%%%%%%%%%%%%
%%%%%%%%%%%%%%%%%%%%%%%%%%%%%%%%%%%%%%%%%%%%%%%
\subsection{Disk Heating}
\label{ssec:heating}
We consider both stellar irradiation and accretion heating and give the temperature $T$ in the disk interior as
\begin{equation}  \label{eq:teq}
	T = \pr{T_{\rm irr}^{4} + T_{\rm acc,\,MHD}^{4}}^{1/4},
\end{equation}
where $T_{\rm irr}$ and $T_{\rm acc,\,MHD}$ represent the contributions from irradiation and accretion, respectively.

The temperature in the irradiation-dominated disk is approximately given by \citep{Kusaka1970Growth-of-Solid,Chiang1997Spectral-Energy}
\begin{eqnarray}
	T_{\rm irr} 
	&=&  110 \pf{r}{1 \au}^{-3/7} \pf{L}{\Lsun}^{2/7} \pf{M}{\Msun}^{-1/7}   \label{eq:int-tirr}~\rm K,
\end{eqnarray}
where $L$ is the stellar luminosity.
\eqref{eq:int-tirr} assumes that stellar light is absorbed at about 4 $H$ above/below the midplane and half of the star is hidden by the innermost region.

\begin{figure}
	\centering
	\includegraphics[width = \hsize,clip]{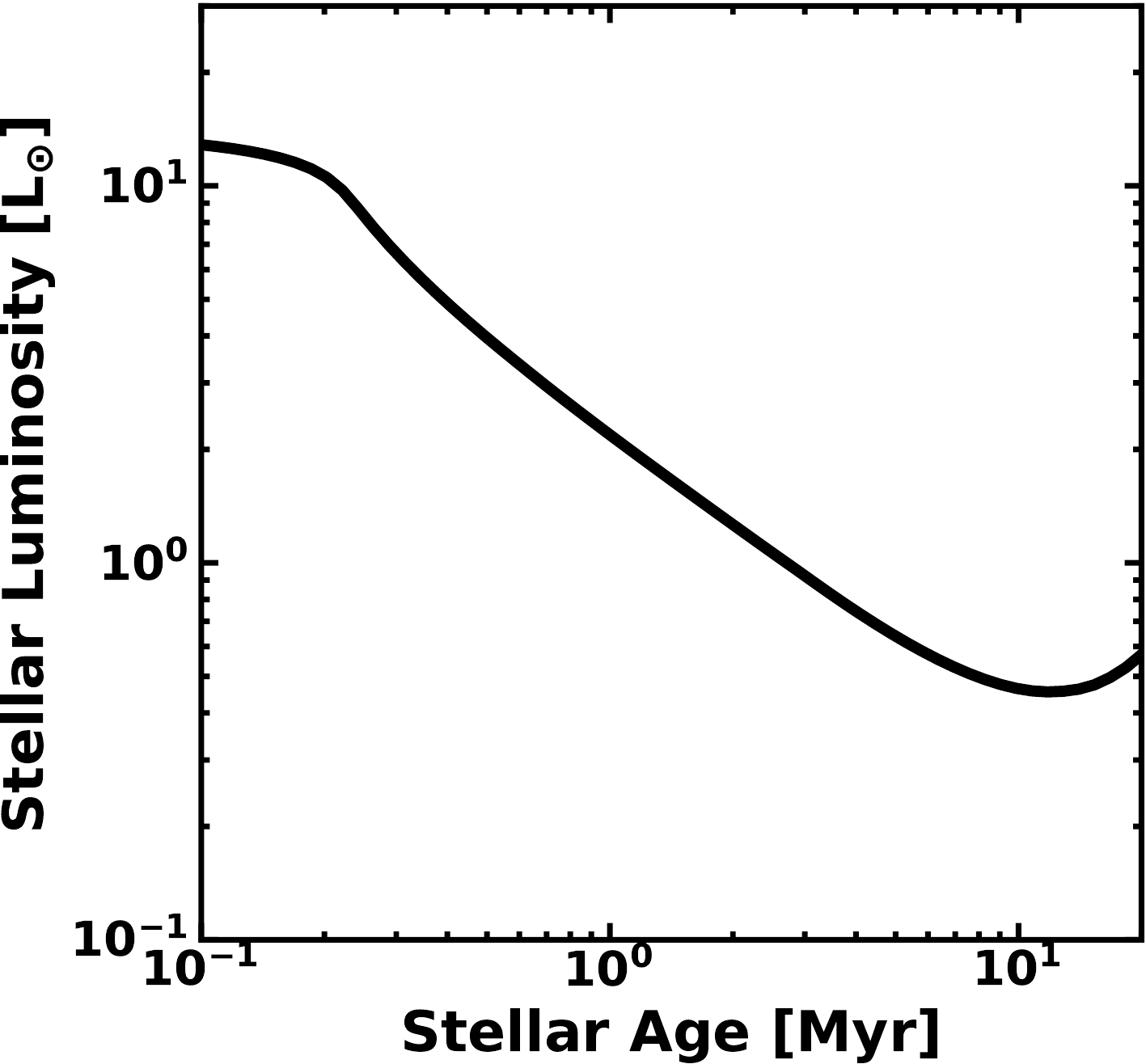}
	\caption{
	Temporal evolutions of stellar luminosity $L$ used in our calculation,
	which are based on \citet{Feiden2016Magnetic-inhibi} (see text for details).
	}
	\label{fig:L}
\end{figure}

We take $L$ to be dependent on $t$ using the evolutionary model for a $1 \Msun$ star by
\citet{Feiden2016Magnetic-inhibi}, as shown in Figure \ref{fig:L}.
We note that, in their model, a $1 \Msun$ star has the deuterium burning from 0.05 to 0.2 Myr and therefore $L$ is kept high.
Here, we neglect the contribution of accretion luminosity to $L$ and assume that the growth of stellar mass can be neglected for simplicity (see \citet{Kimura2016aFrom-birth-to-d} and Section \ref{sec:stel_evol} for the uncertainties in the pre-main-sequence evolution).

For $T_{\rm acc,\,MHD}$ at the midplane, we use (see Appendixes \ref{app:Tderiv} and \ref{app:en-bal} for the derivation)
\begin{equation}\label{eq:T-thick-simple-main}
	T_{\rm acc,\,MHD} = \left[ \pr{ \frac{3 \dot{M} \Omega^{2} f_{\rm heat}}{32 \pi \sigma} } \pr{  \tau_{\rm heat}
					        +    \frac{  1 }{\sqrt{3}} }  \right ]^{1/4} ,
\end{equation}
where $\sigma$ is the Stefan-Boltzmann constant,
and 
$\tau_{\rm heat}$ is an effective optical depth at the midplane and approximately corresponds to the optical depth from infinity to the bottom of the heating layer (see Equation (50) in \citetalias{Mori2019Temperature-Str} and \eqref{eq:tau_heat} for the exact expression). 
The dimensionless parameter $f_{\rm heat}$ ($\leq 1$) is the fraction of the accretion heat deposited inside the disk, 
with $f_{\rm heat} < 1$ when the wind material carries away part of the energy generated by accretion.
We set the default value of $f_{\rm heat}$ to 1 to give a maximum estimate of $T_{\rm acc,\,MHD}$, where $f_{\rm heat} = 1$ is the limiting case that all the liberated energy is consumed for heating.
We discuss the effects of varying $f_{\rm heat}$ in Section \ref{sssec:SL-f}.

We assume that the Rosseland mean opacity $\kappa_{\rm R}$ is constant throughout the disk and write 
\begin{equation} \label{eq:tau-heat}
    \tau_{\rm heat} = \kappa_{\rm R} \Sigmaheat ,
\end{equation}
where $\Sigmaheat$ represents the mass column depth from infinity to the bottom of the heating region (see Appendix \ref{app:Tderiv} and \eqref{eq:Sigmaheat} for the exact expression of $\Sigma_{\rm heat}$).
The remaining task is to evaluate $\Sigmaheat$.

%%%%%%%%%%%%%%%%%%%%%%%%%%%%%%%%%%%%%%%%%%%%%%%
%%%%%%%%%%%%%%%%%%%%%%%%%%%%%%%%%%%%%%%%%%%%%%%
\subsection{Estimating the Heating Layer Depth from Disk Ionization Structure} \label{ssec:heat-depth}

\begin{figure}
	\centering
	\includegraphics[width = \hsize,clip]{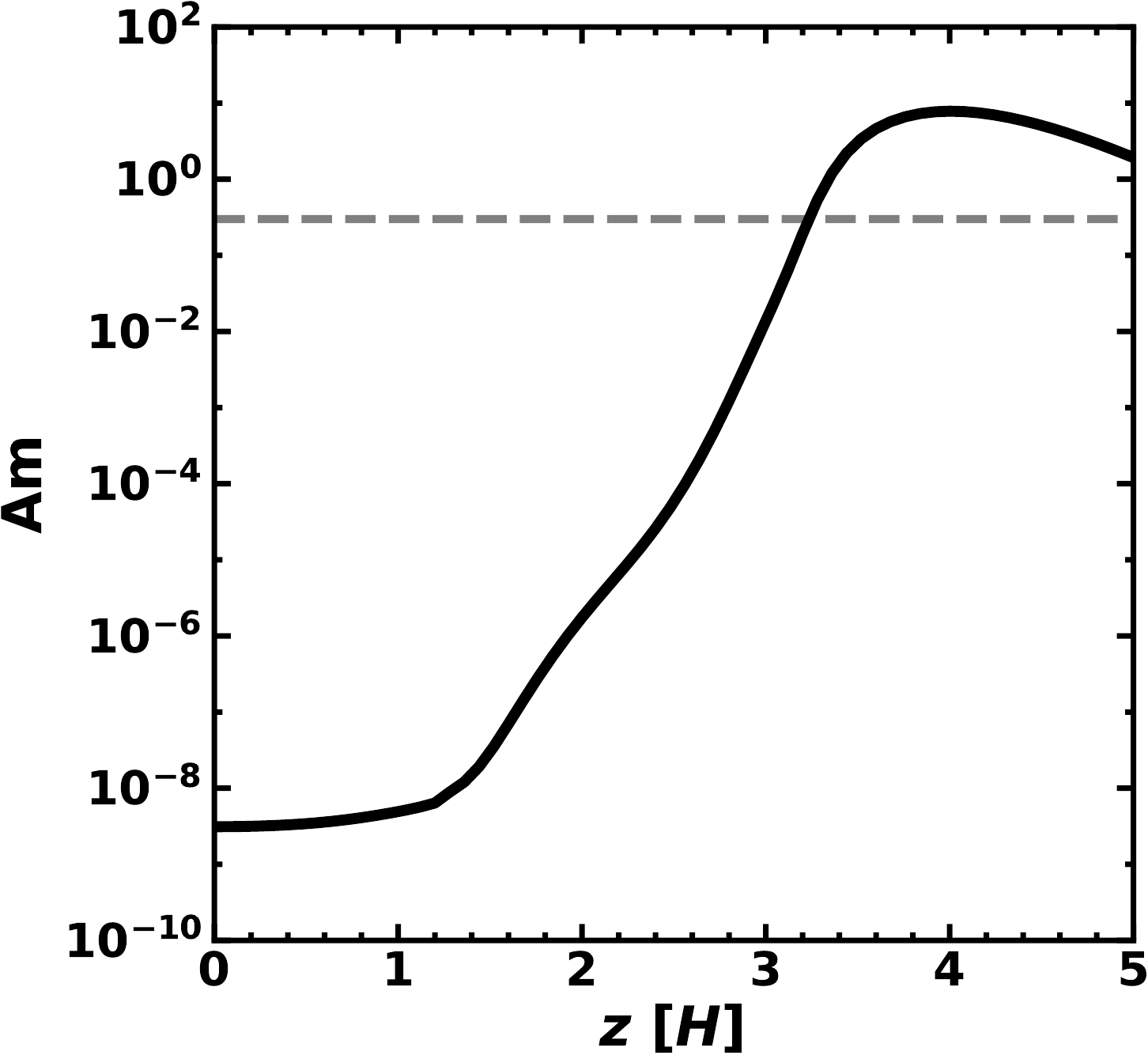}
	\caption{
	Vertical profile of Am at $r=1$ au of the MHD accretion disk model at $t=0.6$ Myr.
	The dashed horizontal line indicates the critical ambipolar Elsasser number (Am = 0.3) of the current layer. 
	}
	\label{fig:Am}
\end{figure}

We estimate $\Sigmaheat$ from the vertical resistivity structure of the disk.
We particularly focus on ambipolar diffusion because it is the dominant non-ideal MHD effect near the disk surface. 
The strength of ambipolar diffusion can be measured by the 
the ambipolar Elsasser number  
\begin{equation}
	\Am = \frac{v_{\rm A}^{2}}{ \etaA \Omega}, 
\end{equation}
where $v_{\rm A}$ is the Alfv\'{e}n speed and $\etaA$ is the ambipolar diffusivity.
The ambipolar Elsasser number is in general independent of the magnetic field strength because both $v_{\rm A}^2$ and $\eta_{\rm A}$ scale quadratically with it.
In inner regions of protoplanetary disks, $\Am$ usually falls below unity at the midplane and increases toward the disk surface \citep{Wardle2007Magnetic-fields,Salmeron2008Magnetorotation,Bai2013aWind-driven-Acc}.
The vertical distribution of Am is shown in Figure \ref{fig:Am}.
MHD simulations show that electric currents decay where $\Am \ll 1$ \citep[][\citetalias{Mori2019Temperature-Str}]{Bai2013aWind-driven-Acc,Gressel2015Global-Simulati}. 
Therefore, one can expect that a current layer should always lie above the critical layer where ${\rm Am}$ is around unity.
In this study, we conservatively take the critical Elsasser number to be $0.3$ and write 
\begin{equation} \label{eq:f-depth}
    \Sigmaheat = f_{\rm depth} \SigmaAm ,
\end{equation}
where the prefactor $f_{\rm depth}$ accounts for the fact that an actual current sheet can occur above the $\Am=0.3$ layer (see Figure \ref{fig:Am}), and $\SigmaAm$ is the gas surface density above the layer.  
In Appendix~\ref{app:z_heat}, we demonstrate that all MHD simulations presented in \citetalias{Mori2019Temperature-Str} satisfy $f_{\rm depth} \la 1$.
The typical ranges of $f_{\rm depth}$, which is affected by Hall effect, are $\sim$ 0.3--1 and $\sim$ 0.01--0.1 when the vertical magnetic field is parallel and anti-parallel to the disk rotation axis, respectively. 
As in the case of $f_{\rm heat}$, we take the default value of $f_{\rm depth}$ to unity to provide a maximum estimate for $T_{\rm acc,\,MHD}$, but also quantify the effect of varying the value of $f_{\rm heat}$ in Section \ref{sssec:SL-f}.

To determine the height of the ${\rm Am} = 0.3$ layer, we compute the vertical profile of $\eta_{\rm A}$ using the ionization model of \citetalias{Mori2019Temperature-Str} that includes charged dust grains (see Section 2.1 of \citetalias{Mori2019Temperature-Str} for details).
The grains regulate the ionization fraction of the disk gas by capturing charged particles at a rate roughly proportional to the total surface area of the grains per unit volume \citep{Bai2009Heat-and-Dust-i,Okuzumi2009Electric-Chargi}.
In this study, we fix the grain size to $0.1~\micron$ and the internal density to $1.4 \gcmcm$, and vary the total grain surface area by changing the dust-to-gas mass ratio $f_{\rm dg}$.
The ionizing sources include galactic cosmic rays \citep{Umebayashi2009Effects-of-Radi}, stellar X-rays \citep{Igea1999X-Ray-Ionizatio,Bai2009Heat-and-Dust-i}, and radionuclides \citep{Umebayashi2009Effects-of-Radi}.
We represent all ion species with a single species by following \citet{Okuzumi2009Electric-Chargi}.
Using the isothermal density profile \eqref{eq:rho}, the height $z_{\rm Am= 0.3}$ of the ${\rm Am= 0.3}$ layer is related to the column depth of the layer as 
\begin{equation}
	\SigmaAm = \frac{\Sigma}{2} {\rm erfc}\pr{\frac{z_{\rm Am=0.3}}{\sqrt{2}H}} .
\end{equation}

%%%%%%%%%%%%%%%%%%%%%%%%%%%%%%%%%%%%%%%%%%%%%%%
\subsection{Numerical Procedures and Parameter Choices}
%%%%%%%%%%%%%%%%%%%%%%%%%%%%%%%%%%%%%%%%%%%%%%%
To summarize, Equations~(\ref{eq:Sigma}) and (\ref{eq:teq}) determine the surface density $\Sigma(r)$ and midplane temperature $T(r)$ of our quasi-steady wind-driven accretion disk model.  Because the right-hand sides of these equations involve gas density and temperature, one solves them iteratively to obtain a self-consistent solution. We set the temperature tolerance to be  $10^{-4}$. 
%We have confirmed that the converged temperature profile does not depend on the profile that is initially set in the iterative calculation.
We have confirmed that the converged temperature profile does not depend on the input profile of the initial guess. 
At each iteration, we also compute the vertical distribution of the ionization fraction to determine the column depth $\Sigmaheat$ of the heating layers given by \eqref{eq:f-depth}. 
When \Am at the midplane is higher than 0.3, we take $\SigmaAm$ to be $\Sigma/2$ 
by assuming that the heating occurs at the midplane.

Our model involves five parameters  
$\{ \kappa_{\rm R}, f_{\rm dg}, \overline{\alpha_{z \phi}}, 
f_{\rm heat}, f_{\rm depth}\}$. 
The default values are taken to be 
$\kappa_{\rm R} = 5 \cmcmg$, $f_{\rm dg} = 0.01$, $\overline{\alpha_{z \phi}} = 10^{-3}$, $f_{\rm heat} = 1$, and $f_{\rm depth} = 1$.
The default value of $\kappa_{\rm R}$ is consistent 
with the opacity for 0.1 $\micron$ sized grains with the ISM abundance computed by \citet{Pollack1985A-calculation-o}.

In this study, we treat $\kappa_{\rm R}$ and $f_{\rm dg}$ as independent parameters, differently in \citetalias{Mori2019Temperature-Str}, to investigate their effects on the disk temperature.
Effectively, we give the total {\it extinction} and {\it geometric} cross sections of dust grains per unit gas mass independently. 
Because the ratio between the two generally depends on the composition and size distribution of grains, one can interpret our treatment as taking into account uncertainties in grain composition and size distribution. 

%%%%%%%%%%%%%%%%%%%%%%%%%%%%%%%%%%%%%%%%%%%%%%%
%%%%%%%%%%%%%%%%%%%%%%%%%%%%%%%%%%%%%%%%%%%%%%%
%%%%%%%%%%%%%%%%%%%%%%%%%%%%%%%%%%%%%%%%%%%%%%%
\section{Migration of the Snow Line}\label{sec:SL}

In this section, we present the evolution of the water snow line location from our MHD disk model and compare it with the prediction from the standard viscous disk model.  
For each disk model, we calculate the radial temperature profile from 0.08 au to 20 au and search for the snow line location assuming the ice sublimation temperature of 170 K.
For the viscous model, we express the viscosity as $\nu = \alpha c_{\rm s} H$ and take the parameter $\alpha$ to be constant throughout the disk.
Assuming steady accretion, the gas surface density of the viscous disk is 
 $\Sigma = \dot{M}/(3\pi \alpha c_{\rm s}H)$ \citep{Lynden-Bell1974The-evolution-o}.
The midplane temperature is given by 
\begin{equation}
    T = \pr{T_{\rm irr}^4 + T_{\rm acc,\,visc}^4 }^{1/4},
\end{equation}
where
\begin{equation}\label{eq:T-visc}
	T_{\rm acc,\,visc} =  \left[ \pr{  \frac{9 \dot{M} \Omega^{2}  }{32 \pi \sigma}  } \pr{ \frac{\tau_{\rm mid}}{2} + \frac{1}{\sqrt{3}}} \right]^{1/4}
	\approx \pr{ \frac{3 \kappa_{\rm R} \dot{M}^2 \Omega^{3}  }{128 \pi^2 \sigma \alpha c_{\rm s}^2 } }^{1/4} 
\end{equation}
represents the contribution from viscous accretion heating with $\tau_{\rm mid} = \kappa_{\rm R}\Sigma/2$ being the vertical optical depth to the midplane \citep[][]{Hubeny1990Vertical-struct,Nakamoto1994Formation-early,Oka2011Evolution-of-Sn}.
The effective optical depth $\tau_{\rm heat}$ for the vertically uniform viscosity is $\tau_{\rm heat} = \tau_{\rm mid}/2$ (\eqref{eq:tauheat-visc}).
The final expression of \eqref{eq:T-visc} assumes $\tau_{\rm heat} \gg 1$, which holds true in our calculations (see Figure~\ref{fig:temp}).  
Note that \eqref{eq:T-visc} should be taken as an implicit equation for $T_{\rm acc,\,visc}$ because the right-hand side depends on $T_{\rm acc,\,visc}$ through $c_{\rm s}^2 \propto T_{\rm acc,\,visc}$.

%For a fair comparison, we take $\alpha$ so that the MHD and viscous disk models have the similar gas surface density.
%Since $\alpha \sim (r/H) \overline{ \alpha_{z \phi}}$ (see the discussion in Section~\ref{sec:Sigma}), 
%for a typical turbulent protoplanetary disk with $H/r \sim 0.1$, one has $\alpha \sim 10\overline{ \alpha_{z \phi}}$. 
%Therefore, we set $\alpha = 10^{-2}$.
%
%For a fair comparison, we take $\alpha$ so that the MHD and viscous disk models have the similar gas surface density.
%Considering the $\dot{M}$--$\Sigma$ relation for the MHD model and viscous model (see Equations \ref{eq:Sigma}),
%$\alpha$ that gives the same value of $\Sigma$ is 
%\begin{equation}
%%	\frac{\alpha}{\alpha_{z\phi}} = 0.5 \pf{T_{\rm MHD}}{T_{\rm visc}}^{1/2}  \pf{r}{H_{\rm visc}} .
%	\frac{\alpha}{\alpha_{z\phi}} = \frac{0.5 c_{\rm s, MHD} r \Omega }{c_{\rm s, visc}^{2}}   .
%\end{equation}
%where $c_{\rm s, MHD}$ and $c_{\rm s, visc}$ are the sound speeds for the temperature in the MHD and viscous models, respectively.
%and are the functions of the disk temperature.
%
%
%
%In this paper, we set  $\alpha = 10^{-2}$ because 
%When we set $\alpha = 10^{-2}$, 
%this fraction is in the range of 5--15
% 
% 
%For a temperature 
%the fraction varies 5--15. 
%Therefore, we set $\alpha = 10^{-2}$ to take $\alpha$ to be $\alpha \sim 10\overline{ \alpha_{z \phi}}$.
%

For a fair comparison, we take $\alpha$ so that the MHD and viscous disk models have the similar gas surface density.
Since $\alpha \sim (r/H) \overline{ \alpha_{z \phi}}$ (see the discussion in Section~\ref{sec:Sigma}), 
for a typical turbulent protoplanetary disk with $H/r \sim 0.1$, one has $\alpha \sim 10\overline{ \alpha_{z \phi}}$. 
Therefore, we set $\alpha = 10^{-2}$.
Although this is just an order-of-magnitude estimate, 
we have confirmed that the ratio of $\Sigma$ in the MHD model to that in the viscous one is in the range of $\approx$ 0.5--1.4 at 1 au for the fiducial parameter set.

%%%%%%%%%%%%%%%%%%%%%%%%%%%%%%%%%%%%%%%%%%%%%%%
%%%%%%%%%%%%%%%%%%%%%%%%%%%%%%%%%%%%%%%%%%%%%%%
\subsection{Fiducial Case}\label{ssec:SL-fid}

\begin{figure*}
	\centering
	\includegraphics[width = .84\hsize,clip]{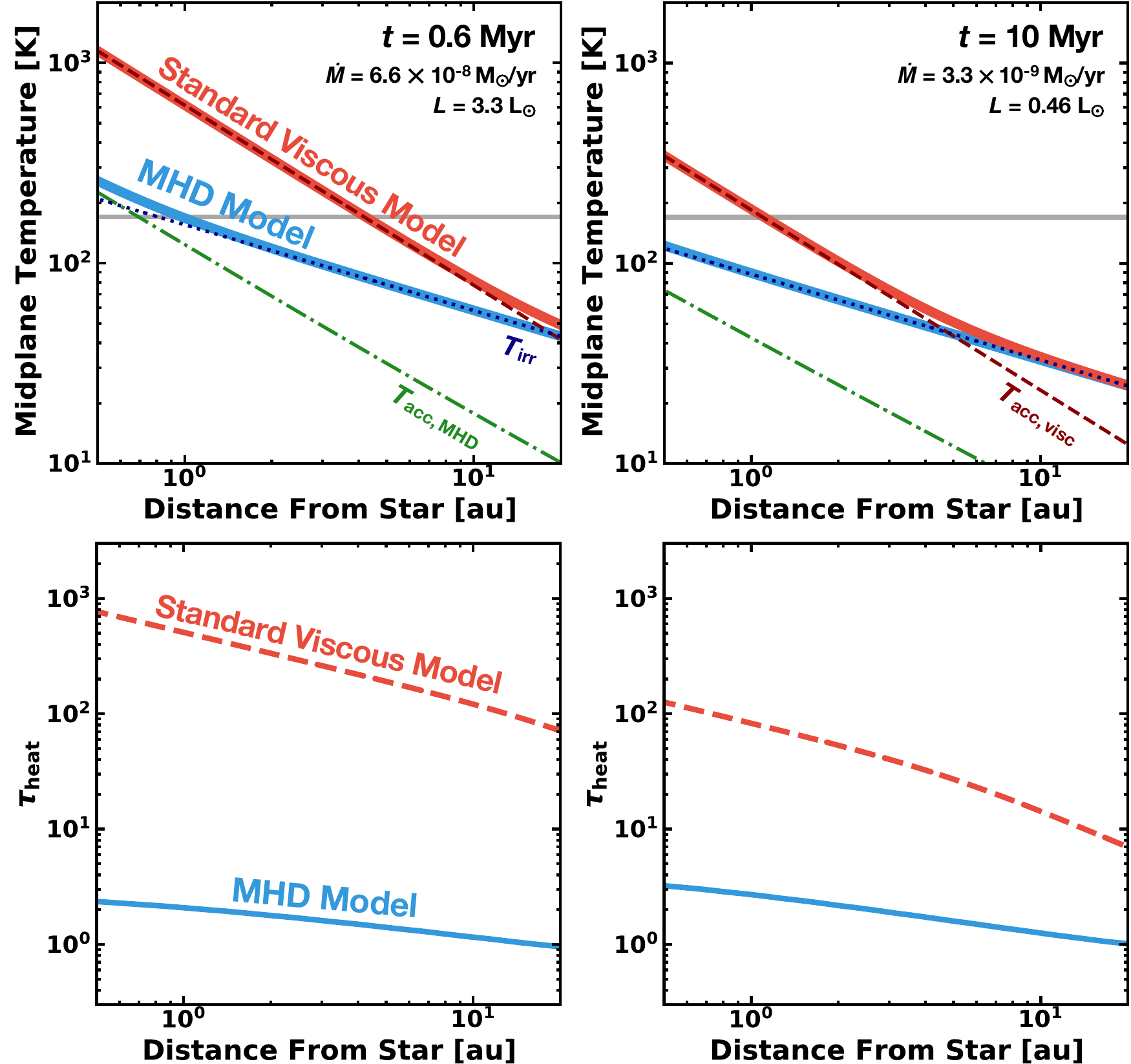}
	\caption{
	Upper panels: radial profiles of the midplane temperature from the MHD accretion disk model (\eqref{eq:teq}; thick solid line) and from the viscous accretion disk model (\eqref{eq:T-visc}; thin solid line) 
	at $t$ = \tcrossMHD Myr (left panel) and $t$ = \tcrossvisc Myr (right panel). 
	For reference, the dotted, dot-dashed, and dashed lines show $T_{\rm irr}$, $T_{\rm acc,\,MHD}$, and $T_{\rm acc,\,visc}$ (Equations (\ref{eq:int-tirr}), (\ref{eq:T-thick-simple-main}), and (\ref{eq:T-visc})), respectively.
	The horizontal solid line marks the sublimation temperature of water ice, $T = 170\,\K$.
	Lower panels: radial profiles of the effective optical depths $\tau_{\rm heat}$ at the midplane for the viscous model ($\tau_{\rm heat} = \tau_{\rm mid}/2$; \eqref{eq:tauheat-visc}; dashed line) and for the MHD model (\eqref{eq:f-depth}; solid line). 
	}
	\label{fig:temp}
\end{figure*}

We begin by the default case where 
$\kappa_{\rm R} = 5 \cmcmg$, $f_{\rm dg} = 0.01$, $\overline{\alpha_{z \phi}} = 10^{-3}$, $f_{\rm heat} = 1$, and $f_{\rm depth} = 1$.
The upper panels of Figure \ref{fig:temp} show the radial temperature profiles from the MHD and viscous disk models at two different stellar ages
(see Figure 4 in \citetalias{Mori2019Temperature-Str} for the typical vertical temperature structure).
In the viscous accretion model, accretion heating is the dominant source of disk heating at $r \la 10~\rm au$ (for $t =$ \tcrossMHD Myr) 
 and $r \la 3~\rm au$  (for $t =$ \tcrossvisc Myr).  
In contrast, in the MHD model, accretion heating only gives a minor contribution, and the midplane temperature is approximately given by $T_{\rm irr}$ at all radii. 
For both models, the disk cools with time, but for different reasons. 
In the viscous model, the temperature decrease is mainly due to the decrease of $\dot{M}$ in $T_{\rm acc,\,visc}$ (\eqref{eq:T-visc}). 
In the MHD model, the temperature evolution is rather driven by the stellar evolution shown in Figure~\ref{fig:L}. 

The inefficient accretion heating in the MHD model can be understood by looking at the optical depth of the heating layer, $\tau_{\rm heat}$, 
shown in the lower panels of Figure \ref{fig:temp}.
In the viscous model, the effective optical depth $\tau_{\rm heat}$ at the midplane is much higher than unity, 
with $\tau_{\rm heat} \approx 500$ (for $t = $ \tcrossMHD Myr) to $\approx 80$ (for $t = $ \tcrossvisc Myr) at 1 au. 
Because of the strong blanketing effect, the accretion heating is the dominant heating mechanism in the inner part of the viscous disk. 
In contrast, in the MHD model, the optical depth $\tau_{\rm heat}$ of the heating layer is only 1--3, and therefore the blanket effect is inefficient. 
This value of $\tau_{\rm heat}$ comes from our ionization calculations showing that 
$\Sigma_{\rm heat} \approx 0.2$--$0.6 \gcmcm$.
Interestingly, the depth of the heating layer is insensitive to the disk age and radial location.
This is because $\Sigma_{\rm heat}$ partly reflects the attenuation depth of ionizing X-rays. 
It should be noted, however, that $\Sigma_{\rm heat}$ also depends on the recombination rate of charged particles and hence on the abundance of small grains in the disk. 
We show this in Section \ref{sssec:SL-fdg}. 

%%%%%%%%%%%%%%%%%%%%%%%%%%%%%%%%%%%%%
\begin{figure}
	\centering
	\includegraphics[width = 0.99 \hsize,clip]{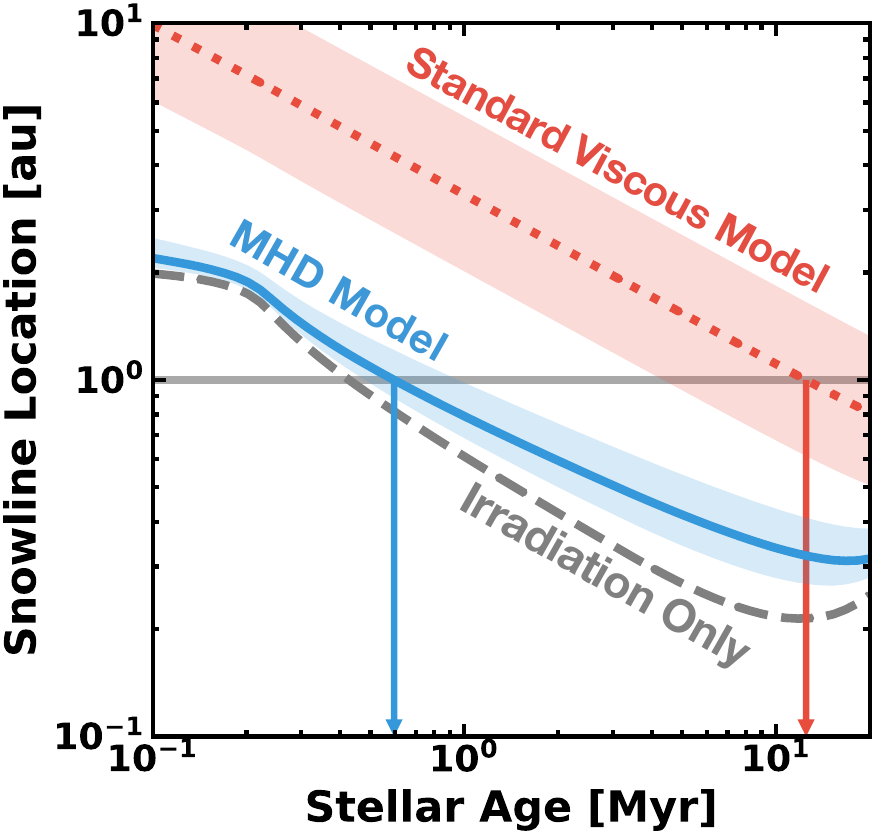}
	\caption{
	Radial location of the snow line at the midplane as a function of the stellar age
	in the MHD model (blue solid) and the standard viscous model (red dotted). 
	The shades show uncertainty coming from observational dispersion of the mass accretion rate \citep[see][]{Hartmann2016Accretion-onto-}.
	The arrows indicate the stellar ages when the snow lines pass 1 au, which are \tcrossMHD and  \tcrossvisc Myr for the MHD and viscous models, respectively. The gray dashed line is for the passive disk with $T = T_{\rm irr}$. 
	}
	\label{fig:sl}
\end{figure}
%%%%%%%%%%%%%%%%%%%%%%%%%%%%%%%%%%%%%
In both disk models, the snow line migrates inward because the disk cools with time. 
Figure \ref{fig:sl} shows the radial location of the snow line at the midplane in the two models as a function of time.
In the viscous disk model, the snow line passes 1 au at $t \approx$ \tcrossvisc Myr.
In contrast, in the MHD model, the snow line arrives 1 au at $t \approx$ \tcrossMHD Myr because of inefficient accretion heating. 
For comparison, we also show in Figure \ref{fig:sl}
the snow line position in a passively irradiated disk with no internal heating, i.e., $T = T_{\rm irr}$. As expected, the snow line location in the MHD disk almost agrees with that in the passively irradiated disk at least until the snow line crosses 1 au. 
 
%%%%%%%%%%%%%%%%%%%%%%%%%%%%%%%%%%%%%%%%%%%%%%%
\subsection{Parameter Dependence}\label{ssec:SL-pdep}
%%%%%%%%%%%%%%%%%%%%%%%%%%%%%%%%%%%%%%%%%%%%%%
\newcommand\pssize{0.99}

We here study the dependence of the snow line tracks in the MHD model on the opacity, ionization fraction, accretion stress, $f_{\rm heat}$, and $f_{\rm depth}$
to understand when the snow line track is largely affected by those parameters. 
The used parameters and obtained results are summarized in Table 1.

\begin{deluxetable*}{lLLLLLLl}[t]
\tablecaption{Summary of used parameters and the time $t_{\rm SL, 1 au}$ that the snow line reaches 1 au.}
\tablehead{
\colhead{Model } &
\dcolhead{ \kappa_{\rm R} ~{\rm [cm^{2} \,g^{-1}]} } & 
\dcolhead{ f_{\rm dg} } & 
\dcolhead{ \overline{\alpha_{z\phi}} } &
\dcolhead{ f_{\rm heat} } &
\dcolhead{ f_{\rm depth} } &
\colhead{ $t_{\rm SL, 1 au}$ [Myr] } &
\colhead{Note}
}
\startdata
MHD     & 5 & 0.01 & 10^{-3} & 1 & 1 & 0.60 & Figure \ref{fig:sl}; fiducial case \\
viscous & 5 & 0.01 & 10^{-3} & 1 & 1 & 12   & Figure \ref{fig:sl}  \\
MHD     & 50 & 0.01 & 10^{-3} & 1 & 1 & 1.8 & Figure \ref{fig:sl-opac}  \\
MHD     & 0.5 & 0.01 & 10^{-3} & 1 & 1 & 0.47 & Figure \ref{fig:sl-opac}  \\
MHD     & 5 & 0.1 & 10^{-3} & 1 & 1 & 0.50 & Figure \ref{fig:sl-fdg}  \\
MHD     & 5 & 0.001 & 10^{-3} & 1 & 1 & 2.2 & Figure \ref{fig:sl-fdg}  \\
MHD     & 5 & 0.001 & 10^{-2} & 1 & 1 & 2.6 & Figure \ref{fig:sl-alppz}  \\
MHD     & 5 & 0.001 & 10^{-4} & 1 & 1 & 2.1 & Figure \ref{fig:sl-alppz}  \\
MHD     & 5 & 0.001 & 10^{-3} & 0.1 & 1 & 0.61 & Figure \ref{fig:sl-fheat} \\
MHD     & 5 & 0.001 & 10^{-3} & 1 & 0.1 & 0.64 &   
\enddata
\end{deluxetable*}

%%%%%%%%%%%%%%%%%%%%%%%%%%%%%%%%%%%%%%%%%%%%%%%
\subsubsection{Variation with the Opacity}\label{sssec:SL-opacity}
%%%%%%%%%%%%%%%%%%%%%%%%%%%%%%%%%%%%%%%%%%%%%%

\begin{figure}
	\centering
	\includegraphics[width = \pssize \hsize,clip]{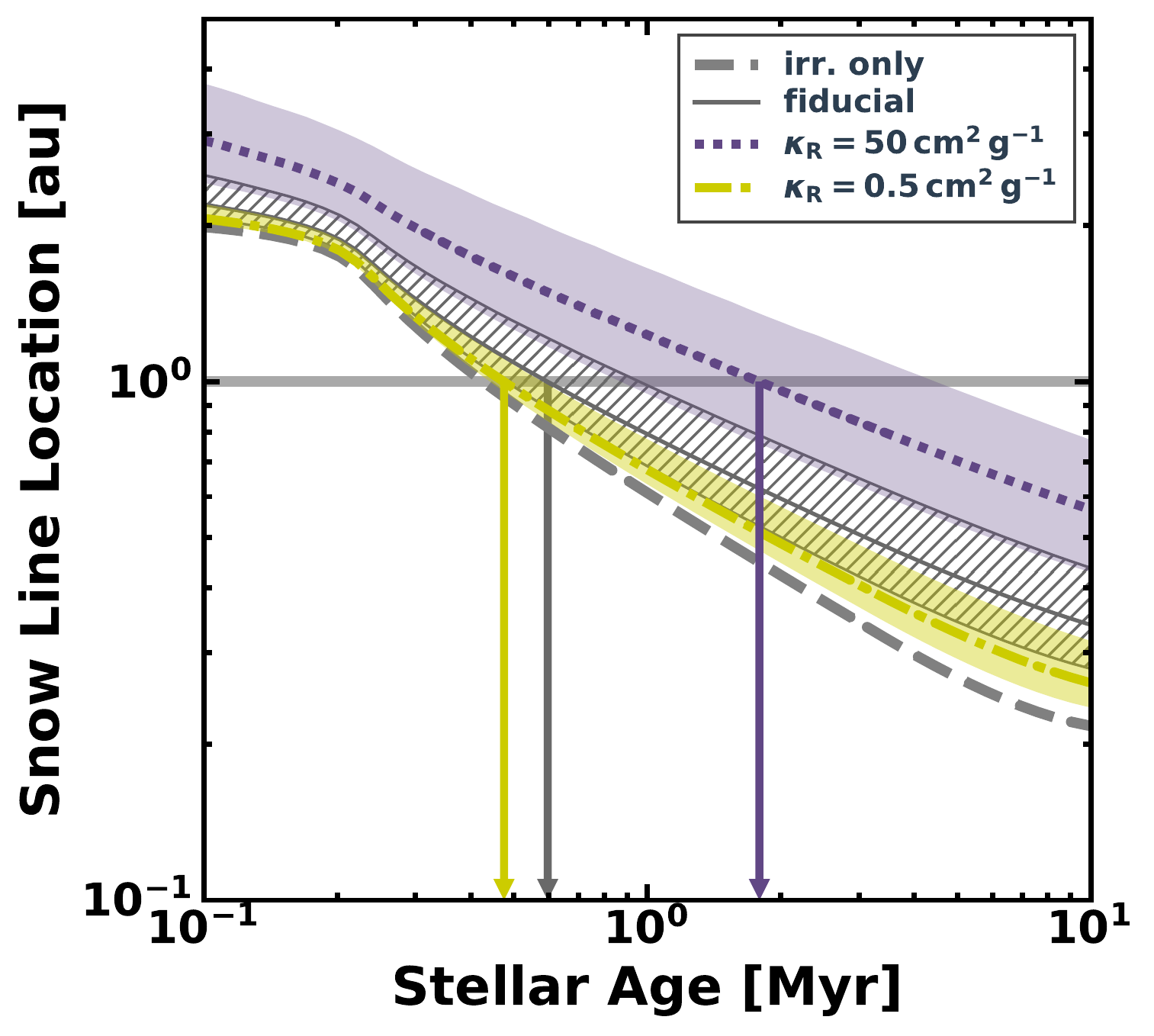}
	\caption{
	Evolution of the snow line location in the MHD model for different values of $\kappa_{\rm R}$. 
	The shades show uncertainty coming from observational dispersion of the mass accretion rate.
	The arrows indicate the stellar age when the snow lines pass 1 au. The gray dashed line is for the passive disk with $T = T_{\rm irr}$. 
	}
	\label{fig:sl-opac}
\end{figure}

Since $\tau_{\rm heat} = \kappa_{\rm R} \Sigma_{\rm heat}$, opacity is an important factor in our model.
Figure \ref{fig:sl-opac} shows the snow line tracks with varying the opacity from the fiducial value of $\kappa_{\rm R} = 5 \cmcmg$. 
When we increase $\kappa_{\rm R}$ to $50 \cmcmg$, $\tau_{\rm heat}$ is increased to $\approx 20$, and consequently the arrival of the snow line at 1 au is delayed to $t \sim$ 2 Myr.
This demonstrates that accretion heating can dominate over irradiation heating even in MHD accretion disks if the opacity is sufficiently high.  
Decreasing the opacity from the fiducial value has little effect on the snow line location because accretion heating is subdominant for such low opacities.
The depth $\Sigma_{\rm heat}$ of the heating layer is insensitive to $\tau_{\rm heat}$ because non-thermal ionization and recombination reactions depend on temperature only weakly. 
In short, we here find that the arrival time of the snow line is $\gg 1$  Myr when $\tau_{\rm heat} \gg 10$.

%%%%%%%%%%%%%%%%%%%%%%%%%%%%%%%%%%%%%%%%%%%%%%%
\subsubsection{Variation with the Ionization Fraction}\label{sssec:SL-fdg}
%%%%%%%%%%%%%%%%%%%%%%%%%%%%%%%%%%%%%%%%%%%%%%

\begin{figure}
	\centering
	\includegraphics[width = \pssize \hsize,clip]{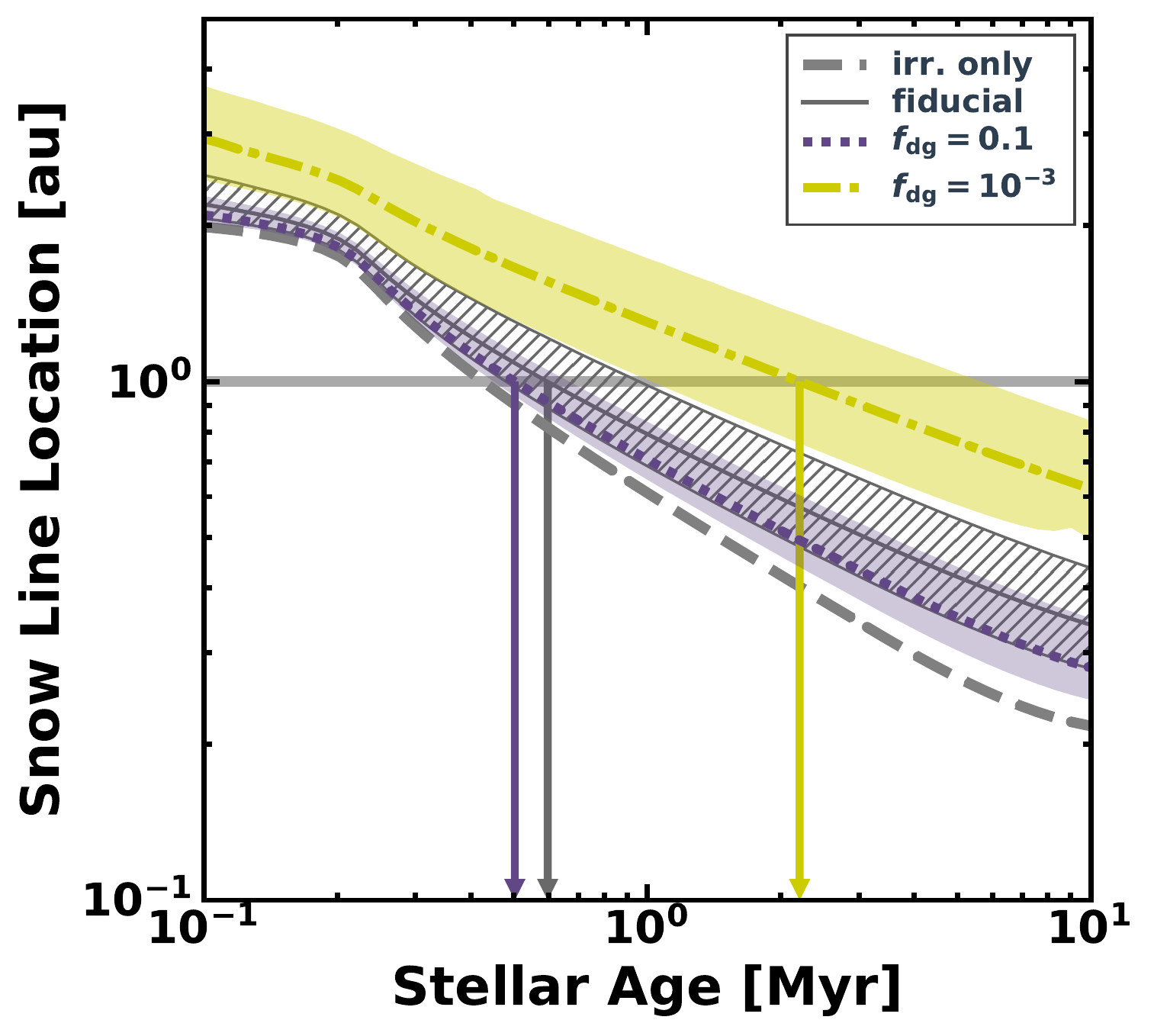}	
	\caption{
	Same as Figure~\ref{fig:sl-opac}, but for different values of $f_{\rm dg}$. Note that $f_{\rm dg}$ is the dust-to-gas mass ratio used in the ionization model and does not affect the disk opacity, which is fixed to $5~\rm cm^{2}~g^{-1}$ here.
	}
	\label{fig:sl-fdg}
\end{figure}

\begin{figure}
	\centering
	\includegraphics[width = \pssize \hsize,clip]{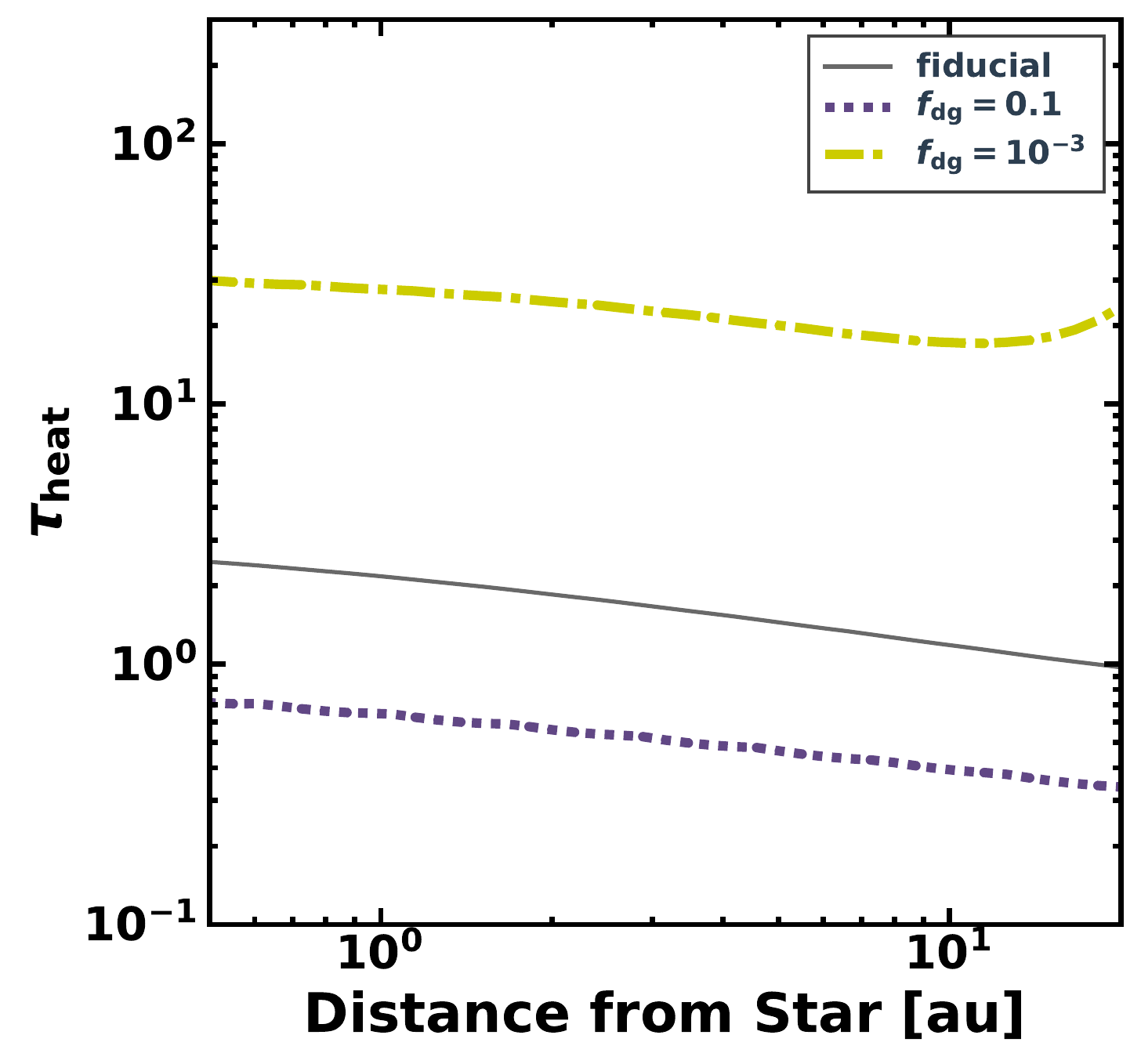}	
	\caption{
	Optical depths $\tau_{\rm heat}$ at the heating layer for different values of $f_{\rm dg}$ at $t = 1$ Myr.
	}
	\label{fig:sl-fdg-odep}
\end{figure}

In our ionization model, the dust-to-gas mass ratio $f_{\rm dg}$ controls the disk ionization fraction  
because grain charging significantly contributes to the removal of ionized particles from the disk gas.
A lower value of $f_{\rm dg}$ gives a higher ionization fraction, a larger $\Sigma_{\rm heat}$, and consequently a higher midplane temperature. 
To see this effect quantitatively, we show in Figure \ref{fig:sl-fdg} the snow line tracks with varying $f_{\rm dg}$ from the fiducial value of $0.01$, with the opacity value fixed.
We also show in Figure \ref{fig:sl-fdg-odep} the radial profile of $\tau_{\rm heat}$ for the cases shown in Figure \ref{fig:sl-fdg}. 
One can see that decreasing $f_{\rm dg}$ by a factor of 10 from the fiducial value leads to an increase of $\tau_{\rm heat}$ by approximately the same factor (corresponding to $\Sigmaheat\approx6 \gcmcm$), resulting in the arrival of the snow line at $1~\rm au$ delayed to 2 Myr. 
Thus, with a sufficiently high ionization fraction, accretion heating can become a dominant heating mechanism even in MHD accretion disks. 
For $f_{\rm dg} \ga 0.01$, accretion heating is subdominant, and hence variation with $f_{\rm dg}$ little affects the snow line evolution.

The optical depth $\tau_{\rm heat}$ depends not only on the opacity but also on the ionization fraction via $\Sigma_{\rm heat}$, and the both in turn depend on the dust model.
For instance, when the total dust abundance is increased, the increase in opacity and the decrease in $\Sigma_{\rm heat}$ can be comparable, and hence $\tau_{\rm heat}$ might not vary significantly.
To properly calculate the disk temperature, it is important to consistently give the opacity and ionization fraction using an identical dust model.

%%%%%%%%%%%%%%%%%%%%%%%%%%%%%%%%%%%%%%%%%%%%%%%
%%%%%%%%%%%%%%%%%%%%%%%%%%%%%%%%%%%%%%%%%%%%%%%
\subsubsection{Variation with the MHD Parameters}\label{sssec:SL-Maxwell}\label{sssec:SL-f}

%\subsubsection{Variation with the Wind Accretion Stress}\label{sssec:SL-Maxwell}
%% COMMENT FROM MK: 
%% Very minor: The first paragraph of Sect. 3.2.3 is a bit strange. I would merge Sects. 3.2.3 and 3.2.4 and its title would be "variation with the MHD parameters"

\begin{figure}
	\centering
	\includegraphics[width = \pssize \hsize,clip]{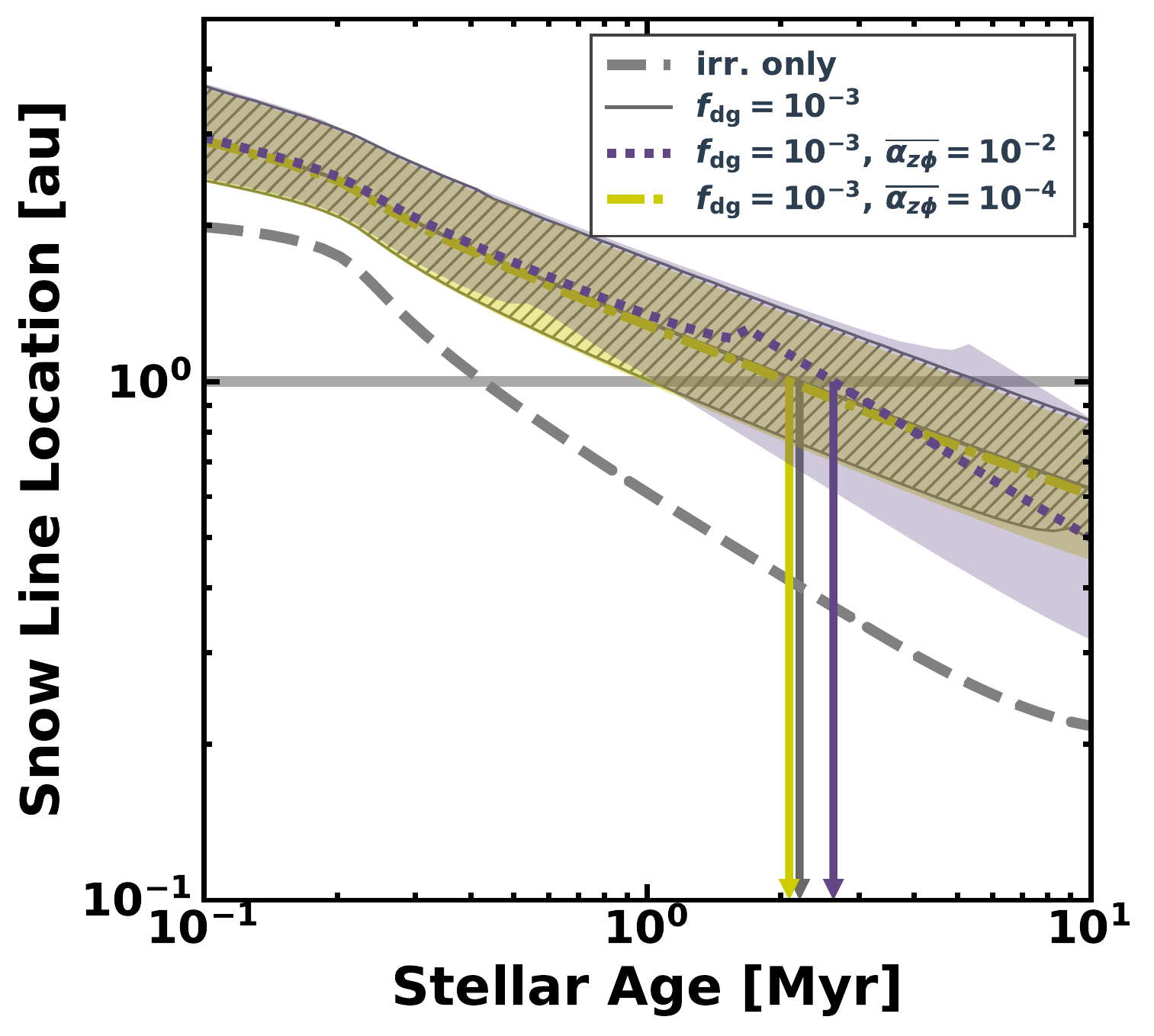}	
	\caption{
	Same as Figure~\ref{fig:sl-opac}, but for $f_{\rm dg} = 10^{-3}$ with different values of $\overline{\alpha_{z \phi}}$.
	The gray line is for the fiducial value of $\overline{\alpha_{z \phi}} = 10^{-3}$. 
	}
	\label{fig:sl-alppz}
\end{figure}

We have shown in Section~\ref{sssec:SL-fdg} that a high $\kappa_{\rm R}$ and/or a low $f_{\rm dg}$ makes accretion heating dominant even in the MHD model. 
The question is now how strongly the level of the accretion heating depends on the rather unconstrained parameters involved in the MHD disk model, namely $\alppz$, $f_{\rm heat}$, and $f_{\rm depth}$.

The magnetic stress $\alppz$ mainly depends on the net flux of the vertical magnetic field \citep[e.g.,][]{Hawley1995Local-Three-dim,Bai2013bWind-driven-Acc} which evolves with the disk evolution \citep[e.g.,][]{Lubow1994Magnetic-field-,Guilet2014Global-evolutio,Okuzumi2014Radial-Transpor,Takeuchi2014Radial-Transpor,Leung2019Local-semi-anal}.
%The radial transport of the magnetic flux also depends on the disk ionization structure \citep[e.g.,][]{Leung2019Local-semi-anal}.
Thus, $\alppz$ may vary widely.

It can be readily expected that the midplane temperature in the MHD model must be insensitive to $\alppz$. 
This parameter determines the magnitude of the wind accretion stress (see \eqref{eq:w_wind}) and controls the disk surface density $\Sigma$ (see \eqref{eq:Sigma}). 
However, it is the column depth $\Sigma_{\rm heat}$ of the heating layer, not the total column density $\Sigma$, that determines $T_{\rm acc,\,MHD}$.
The depth of the heating layer is determined by the ionization structure well above the midplane, and is therefore insensitive to $\Sigma$. 
We demonstrate this in Figure \ref{fig:sl-alppz}, 
where we plot the snow line tracks for $f_{\rm dg} = 10^{-3}$ (see also Figure \ref{fig:sl-fdg}) but with different values of $\alppz$.
It is clear that changing $\alppz$ has essentially no effect on the snow line location.

%%%%%%%%%%%%%%%%%%%%%%%%%%%%%%%%%%%%%%%%%%%%%%%
%%%%%%%%%%%%%%%%%%%%%%%%%%%%%%%%%%%%%%%%%%%%%%%
%\subsubsection{Variation with $f_{\rm heat}$ and $f_{\rm depth}$} \label{sssec:SL-f}

\begin{figure}
	\centering
	\includegraphics[width = \pssize \hsize,clip]{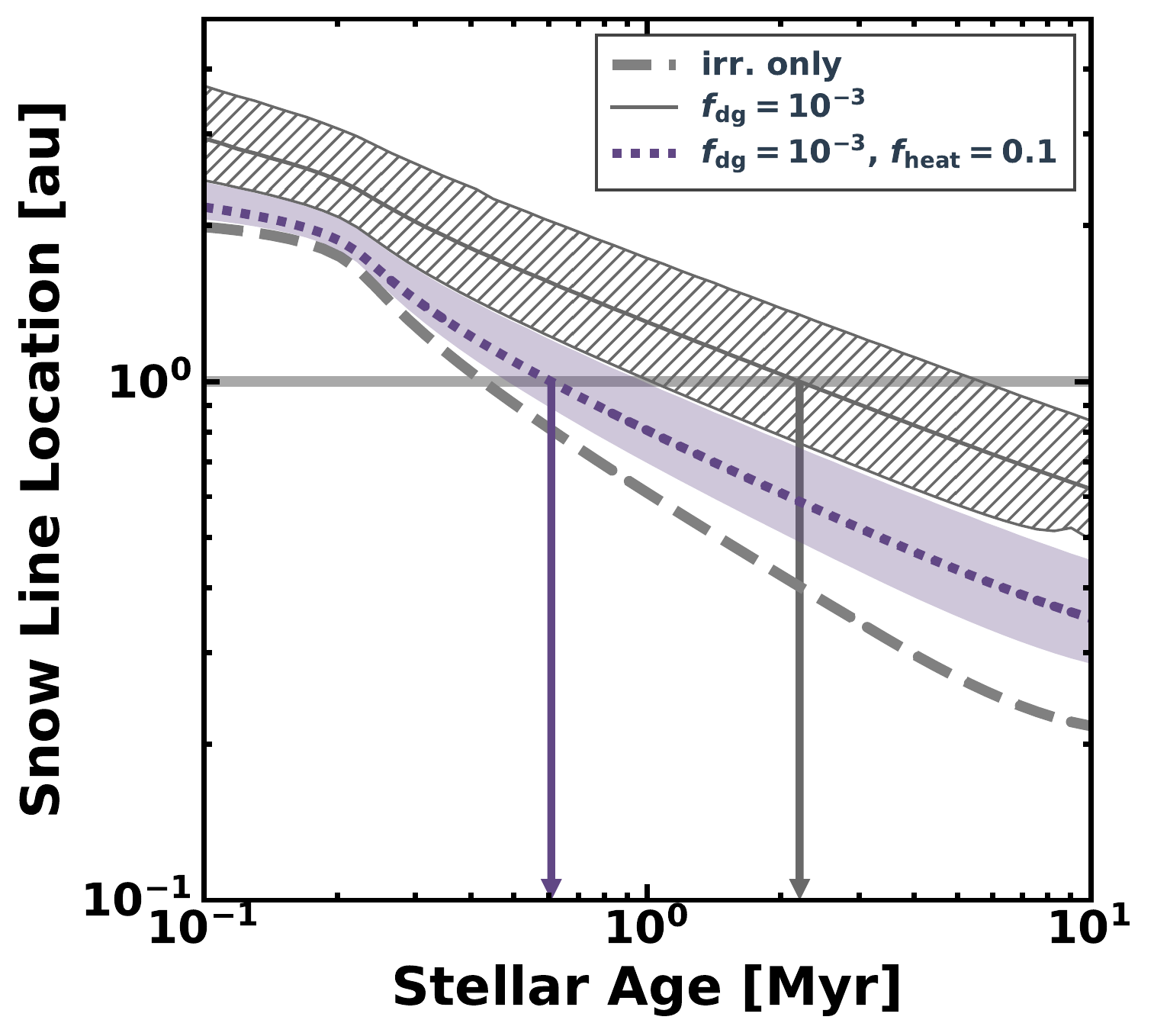}	
	\caption{
	Same as Figure~\ref{fig:sl-opac}, but for $f_{\rm dg} = 10^{-3}$ with different values of $f_{\rm heat}$.  The gray line is for the fiducial value of $f_{\rm heat} = 1$. 
	}
	\label{fig:sl-fheat}
\end{figure}

The calculations shown above adopt $f_{\rm heat} = f_{\rm depth} = 1$.  
Such calculations give the maximum estimate of $T_{\rm acc,\,MHD}$. 
We also vary the two parameters to illustrate how the choice affects the snow line location from our model.

As explained in Sections~\ref{ssec:heating} and Appendix~\ref{app:en-bal}, 
$f_{\rm heat}$ is 
%the fraction of the heat deposited inside the disk in the energy liberated by wind-driven accretion.
the fraction of the energy dissipated inside the disk to the energy liberated by wind-driven accretion.
%$f_{\rm heat}$ is the fraction of the energy liberated by wind-driven accretion.
This number is much smaller than one when the wind carries away a large fraction of the accretion energy.
The value of $f_{\rm heat}$ depends on the polarity of the net vertical magnetic field as well as $f_{\rm depth}$, and the range is approximately 10$^{-3}$ to 1.
In Figure \ref{fig:sl-fheat}, 
we plot the snow line track for $f_{\rm dg} = 10^{-3}$, but now assuming $f_{\rm heat} = 0.1$, i.e., only 10\% of the accretion energy is consumed by the disk's internal heating. 
For such as low value of $f_{\rm heat}$, accretion heating is no longer significant even with $f_{\rm dg} = 10^{-3}$, and the snow line arrives at 1 au within 1 Myr as in the fiducial model (Figure~\ref{fig:sl}).

The factor $f_{\rm depth}$ is less than unity when Joule heating occurs at an altitude higher than where ${\rm Am} = 0.3$.
For instance, as shown in Figure \ref{fig:mod-vs-res} of Appendix \ref{app:z_heat}, 
$f_{\rm depth}$ can be as low as 0.01--0.1 when the magnetic field threading the disk is anti-aligned with the disk rotation axis.
Since $T_{\rm acc,\,MHD} \propto (f_{\rm heat}f_{\rm depth})^{1/4}$ as long as $\tau_{\rm heat} \ga 1$, varying $f_{\rm depth}$ from 1 to 0.1 gives the same effect as varying $f_{\rm heat}$ from 1 to 0.1 (see Figure~\ref{fig:sl-fheat}).

%%%%%%%%%%%%%%%%%%%%%%%%%%%%%%%%%%%%%%%%%%%%%%%
%%%%%%%%%%%%%%%%%%%%%%%%%%%%%%%%%%%%%%%%%%%%%%%
%%%%%%%%%%%%%%%%%%%%%%%%%%%%%%%%%%%%%%%%%%%%%%%
\section{Discussion}\label{sec:discuss}

\subsection{Comparison with \citet{Bethune2020Electric-heatin}} \label{ssec:BL20}

Here we compare our results with \citet{Bethune2020Electric-heatin}, who also investigated accretion heating in magnetic laminar PPDs with non-ideal MHD calculations.
\citet{Bethune2020Electric-heatin} showed that Joule heating can efficiently warm the disk interior, apparently in contradiction with the results by  \citetalias{Mori2019Temperature-Str}.
\citet{Bethune2020Electric-heatin} speculated that the discrepancy arises from the different opacities adopted in the two studies:  
\citet{Bethune2020Electric-heatin} used the opacity of the order of $5 \cmcmg$, while \citetalias{Mori2019Temperature-Str} adopted a ten times smaller value.

The results presented in this study indicate that the combination of the disk opacity and ionization state is a key to understanding the discrepancy. 
According to \eqref{eq:T-thick-simple-main}, 
the temperature of the disk interior is determined by the optical depth $\tau_{\rm heat}$ of the heating layer, which is the product of the opacity and the column depth $\Sigma_{\rm heat}$ of the layer. 
The disk ionization state plays an important role here because it determines $\Sigma_{\rm heat}$. 
As demonstrated in Section~\ref{ssec:SL-fid}, the value of $\kappa_{\rm R} = 5~\rm cm^2~s^{-1}$ alone is not sufficient to warm the disk interior.
Efficient heating also requires a low dust abundance producing a high ionization fraction and hence a high $\Sigma_{\rm heat}$ (Section~\ref{sssec:SL-fdg}). 
Indeed, the fiducial model of \citet{Bethune2020Electric-heatin} adopted a relatively high ionization fraction corresponding to the dust-free limit in our model.
In contrast, in the model of \citetalias{Mori2019Temperature-Str}, the opacity was scaled with the dust-to-gas ratio used in the ionization calculations and therefore decreased whenever $\Sigma_{\rm heat}$ is increased.
These results indicate that treating the opacity and ionization fraction consistently using an identical dust model is crucial for properly evaluating the efficiency of Joule heating.

%%%%%%%%%%%%%%%%%%%%%%%%%%%%%%%%%%%%%%%%%%%%%%%
%%%%%%%%%%%%%%%%%%%%%%%%%%%%%%%%%%%%%%%%%%%%%%%

\subsection{Implications for the Formation of the Inner Solar System} 

%\subsection{Implications for the Formation of the dry Earth} \label{ssec:scenarios}
\subsubsection{Formation of the Dry Earth} \label{ssec:scenarios}

The results presented in Section \ref{sec:SL} show that the snow line in a magnetically accreting disk around a Sun-like star passes the current Earth's orbit as early as 1 Myr after star formation
except for particular parameter choices.
In such an early stage of disk evolution, the region outside the snow line is always abundant in icy particles, and therefore protoplanets at 1 au would acquire a significant amount of water ($\ga 1~\%$ of the protoplanet mass) by accreting the icy particles \citep{Sato2016On-the-water-de,Ida2019aWater-delivery-}.
Therefore, our results strongly indicate that either our Earth did not form at its current orbit or some mechanism prevented icy particles from migrating into the 1 au region in the solar nebula.

There are some possible scenarios that may explain the low water content of the Earth.
One is that Jupiter's core formed earlier than the Earth and carved a gas gap that blocked inward migrating icy particles \citep{Morbidelli2016Fossilized-cond}.
If this is the case, our results presented in Section \ref{sec:SL} suggest that the proto-Jupiter formed within 1 Myr after the solar nebula formation.
This is consistent with what is inferred from a recent meteoritic analysis \citep{Kruijer2017Age-of-Jupiter-}.

Alternatively, \citet{Johansen2021A-pebble-accret} proposed a possibility that when the temperature of planetary atmosphere is high enough, 
the water vapor sublimated from icy pebbles accreting to the protoplanet is recycled back to the PPD. 
This reduces the water content as the protoplanet grows with the pebble accretion, and the final water content might match with the current water content of the Earth. 
%This reduces the water content as the protoplanet grows with the pebble accretion, and the final water content might match with the current value. 
%If this is the case, even for the MHD accretion disk, the final water content of the Earth might be low. 

Another scenario is that the Earth's embryo formed at a close-in orbit ($\sim 0.1$ au) and then migrated to the current orbit after the icy particles in the nebula had been depleted. 
For instance, there are models suggesting that rocky planetesimals form at the inner boundary of the dead zone where the temperature is $\approx 1000$ K \citep[e.g.,][]{Kretke2009Assembling-the-,Dzyurkevich2010Trapping-solids,Drazkowska2013Planetesimal-fo}. 
\citet{Ogihara2015Formation-of-te} show that planetesimals forming in the close-in region move outward when magnetorotationally driven disk winds create a positive surface density slope. 
However, whether the migration of the close-in planetesimals occurs after the icy particles outside the snow line have been sufficiently depleted would depend on the efficiency of the wind-driven accretion and wind mass loss in the inner region.

%\subsection{Isotopic Dichotomy of Planetesimals} \label{ssec:scenarios}
\subsubsection{Dichotomy of Planetesimals} \label{ssec:pldicho}

The position of the snow line may be a key to understanding a dichotomy between carbonaceous chondrites (CCs) and noncarbonaceous chondrites (NCs) in the solar system.
%Parent bodies for CCs and NCs are different in the formation time of their iron core \citep{Kruijer2014Protracted-core}.
Meteoritic analyses of CCs and NCs based on the isotopic compositions suggest that 
the parent bodies were born in the spatially separated reservoirs of CC/NC-like materials
and the reservoirs were formed at different timings \citep[see][]{Kruijer2017Age-of-Jupiter-,Kleine2020The-Non-carbona}.
%
%are different in isotopic com@
%
%Parent bodies for CCs and NCs are suggested to be different in iron core formation time, aqueous alteration time, and isotopic compositions \citep{Kruijer2014Protracted-core}.
\citet{Kruijer2017Age-of-Jupiter-} suggested that 
the two reservoirs were separated due to the early forming Jupiter creating a gap in the solar nebula.
%the gap due to the early forming Jupiter created the two populations of planetesimals which were the reservoirs. 
\citet{Lichtenberg2021Bifurcation-of-}  
proposed another scenario that explains the dichotomy in a viscously evolving disk model.
In their scenario, 
the snow line moves outward as the viscously heated region expands with disk formation during the Class I phase.
Thus, two types of planetesimals with different formation regions and timings are formed around the snow line \citep{Drc-azkowska2018Planetesimal-fo} and are responsible for the dichotomy.

The inefficient accretion heating in the MHD disk model may affect the scenario in \citet{Lichtenberg2021Bifurcation-of-}. 
In the MHD model, at least in the Class II phase, the snow line lies further closer to the central star than assumed in \citet{Lichtenberg2021Bifurcation-of-}.
For the Class I phase, it is uncertain if this is the case,
e.g., a self-gravitational instability might heat the disk as much as viscous heating (see Section \ref{sssec:gi}). 
If accretion heating is inefficient only in the Class II phase, 
the formation region of planetesimals in the Class II phase should overlap that in the Class I phase. 
In this case, numerous planetesimals with compositions intermediate between CCs and NCs may be formed, inconsistent with the meteorite analyses.
If accretion heating is inefficient in both the Class I and Class II phases, the two types of planetesimals could be spatially well separated, 
although they would form in a more inner region than expected in \citet{Lichtenberg2021Bifurcation-of-}.

%%%%%%%%%%%%%%%%%%%%%%%%%%%%%%%%%%%%%%%%%%%%%%%
%%%%%%%%%%%%%%%%%%%%%%%%%%%%%%%%%%%%%%%%%%%%%%%
\subsection{Outward Migration of the Snow Line}\label{sec:outmig}

In this paper, the direction of the snow line migration is only inward as in Figure \ref{fig:sl}, 
whereas in \citet{Oka2011Evolution-of-Sn}, the snow line moves outward after the inward migration.
The outward migration of the snow line occurs when the disk becomes optically thin and thereby the stellar irradiation gets efficient. 
According to \citet{Oka2011Evolution-of-Sn}, the outward migration occurs when the accretion rate decreases to less than $\dot{M} \lesssim 10^{-9} \Mpyr$.
At such low accretion rate, the disk is near dispersal (in our model more than 10 Myrs).
We are primarily concerned with snow line evolution in the bulk disk lifetime, and so the outward migration of the snow line is not focused on.

%%%%%%%%%%%%%%%%%%%%%%%%%%%%%%%%%%%%%%%%%%%%%%%
%%%%%%%%%%%%%%%%%%%%%%%%%%%%%%%%%%%%%%%%%%%%%%%
\subsection{Uncertainty of Stellar Luminosity Evolution}\label{sec:stel_evol}

We showed in Section \ref{ssec:SL-fid} that the snow line evolution in magnetically accreting PPDs is primarily determined by the evolution of the stellar luminosity.
In this study, we have adopted a conventional pre-main-sequence evolutionary model.

We adopt the evolutionary model by \citet{Feiden2016Magnetic-inhibi}, which is also used by \citet{Hartmann2016Accretion-onto-}  to determine stellar mass and age. 
However, we note that the initial condition of \citet{Feiden2016Magnetic-inhibi} is much more luminous than the birthline by \citet{Stahler2005The-Formation-o} and therefore our stellar luminosity and thus $T_{\rm irr}$ are overestimated in particular in the early phase \citep[$\lesssim1\,$Myr; see also Section 2.2 of][]{Hartmann1998Accretion-and-t}.
If time is defined as the time after the luminosity reaches that of the birthline of \citet{Stahler2005The-Formation-o}, 
the time when the snow line passes 1 au becomes earlier to $t \approx 0.2$ Myr for the fiducial case of the MHD model.

The initial luminosity of pre-main-sequence stars may be even lower than that of \citet{Stahler2005The-Formation-o}.
They derived the birthline by assuming a large fraction of accretion energy is injected into the protostar (called high-entropy accretion).
However, recent studies have shown that the evolution of young stars depends on how much entropy of accreting materials is injected into the star 
\citep[e.g.][]{Hartmann1997Disk-Accretion-,Hosokawa2011On-the-Reliabil,Baraffe2012Observed-Lumino,Kunitomo2017Revisiting-the-}.
If the gas looses most of the entropy before it reaches the star (called low-entropy accretion; i.e., most of accretion energy is radiated away),
the luminosity of pre-main-sequence stars becomes much lower than fiducial models.
In this case, the snow line passes 1 au much earlier.

We also note that in the present paper, we neglect the growth of stellar mass and the accretion luminosity, $L_{\rm acc}$\footnote{The accretion luminosity is the radiation from the accretion shock surface, while the intrinsic luminosity comes from the stellar photosphere.} (see Section\,\ref{ssec:heating}).
The accretion still remains vigorous until $\approx 1\,$Myr (see Figure\,\ref{fig:L}).
However, in the case of high-entropy accretion,
$L_{\rm acc}\lesssim1\,{\rm L_\odot}$ after the star leaves the birthline and always satisfies $L_{\rm acc}\ll L_{\rm int}$, where $L_{\rm int}$ is the stellar intrinsic luminosity.
Hence we can safely neglect $L_{\rm acc}$.
In the case of low-entropy accretion, $L_{\rm acc}$ can be higher than $L_{\rm int}$, but the total luminosity is comparable to or less than that of the high-entropy accretion case.
Therefore the stellar luminosity in this paper is unlikely to be underestimated and the conclusion of this paper (i.e., early arrival of the snow line to the present terrestrial-planet orbits) should not be affected by the uncertainties of stellar evolution models.
Future work on this issue should include the protostellar accretion to the stellar evolution model and investigate the influence of the uncertainty in the stellar evolution model on the snow line location.

%%%%%%%%%%%%%%%%%%%%%%%%%%%%%%%%%%%%%%%%%%%%%%%
%%%%%%%%%%%%%%%%%%%%%%%%%%%%%%%%%%%%%%%%%%%%%%%
\subsection{Other Heating Mechanisms} \label{ssec:otherheat}

We here discuss the possibility that other heating mechanisms influence the location of the snow line. 

%%%%%%%%%%%%%%%%%%%%%%%%%%%%%%%%%%%%%%%%%%%%%%%
\subsubsection{Hydrodynamic Turbulence} \label{sssec:hydro}

Hydrodynamic turbulence can be a heat source other than Joule heating.
Hydrodynamic instabilities may generate the turbulence when MRI is fully suppressed \citep{Klahr2018Instabilities-a,Lyra2019The-initial-con,Cui2020Global-Simulati}.
If part of the accretion energy is converted into heat in optically-thick regions, the disk temperature can be increased by the blanketing effect.
It is important to note that even if the accretion is driven by the disk wind rather than the turbulence, the heating by the turbulence can still warm the disk up.

There are at least two conditions for the heating to affect the disk temperature.
The first is the growth of hydrodynamic instabilities in irradiated disks suggested by the MHD model.
\citet{Pfeil2019Mapping-the-Con} investigated the unstable regions for some linear hydrodynamic instabilities (i.e., vertical convective instability, vertical shear instability, and convective overstability).
% The convective instabilities grow at $\approx$ 1 au on the midplane in viscously heated disks,
% but are stable in irradiated disks (see the lowest $\alpha$ case in their Figure 10).
The disk midplane is stable to convective overstability and vertical convection at 1 au in the absence of viscous heating (see the lowest $\alpha$ case in their Figure 10).
%This is because the instabilities of the inner region develop in the temperature structure due to viscous heating.
In addition, the vertical shear instability does not grow in the optically thick region.
Therefore, the linear instabilities may not be a heating source in the optically thick inner region of MHD accretion disks.
On the other hand, some nonlinear instabilities (e.g., zombie vortex instability and subcritical baroclinic instability) may grow even in irradiated disks \citep{Marcus2015Zombie-Vortex-I,Lesur2016On-the-survival,Lyra2019Planet-formatio,Pfeil2019Mapping-the-Con}.
The nonlinear instabilities may drive turbulence if some mechanisms provide sufficient amplitudes.

The second condition is sufficiently strong turbulence around the midplane to affect the disk temperature compared to irradiation.
Hydrodynamic simulations have shown that the Shakura-Sunyaev $\alpha$ parameter showing the turbulence strength averaged over the disk height is $\sim 10^{-4}$--$10^{-3}$
\citep[e.g.,][]{Lyra2014Convective-Over,Stoll2014Vertical-shear-,Marcus2015Zombie-Vortex-I,Flock2020Gas-and-dust-dy}.
Assuming the presence of the uniform turbulent viscosity in the MHD accretion disk, the $\alpha$ parameter of $\approx 3 \times10^{-4}$ is sufficient to affect the disk temperature profile around 1 au at 1 Myr.
However, it is still unclear whether the turbulent viscosity is uniform and enough heat is released inside the disk. 
%The strength of hydrodynamic turbulence would be sufficient for this condition,  

We should note that once any turbulence changes the temperature structure, the turbulence may remain by sustaining the unstable thermal structure.
Instability growing in the upper region could bring the turbulence near the midplane \citep{Nelson2013Linear-and-non-,Klahr2014Convective-Over}. 
Such turbulence may also transport the energy to around the midplane. 
Further numerical studies are necessary for understanding practical criteria for the disk heating.

To summarize this section, 
hydrodynamic turbulence is a potential disk heating source
but it is still unclear whether such turbulence develops and whether it releases enough heat around the midplane.
%The heating delays the passage of the snow line through the terrestrial-planet forming region.

%%%%%%%%%%%%%%%%%%%%%%%%%%%%%%%%%%%%%%%%%%%%%%%

\subsubsection{Shock Heating of Waves Induced by Gravitational Instability}\label{sssec:gi}

Disks at early stages may be subject to the gravitational instability.
Shock heating of waves induced by self-gravitational instability can also warm up the inner disk region \citep[see][]{Boss2005Chondrule-formi,Martin2012On-the-evolutio,Rafikov2016Protoplanetary-}.
This instability typically occurs in the outer disk region, but shock waves excited there may propagate to the inner region \citep{Rafikov2016Protoplanetary-}.
The shock waves release heat around the midplane \citep{Hirose2017Gravito-turbule}, and thus can warm up the disk by the blanketing effect.

However, the unstable region at 0.6 Myr is uncertain but is generally expected to be well beyond $\sim$10 au \citep[see][]{Kratter2016aGravitational-I}, which would be too far away for the wave to propagate to $\sim$ 1 au.
Thus, the gravitational instability would not drive the disk heating when the snow line passes the rocky-planet-forming region.

%%%%%%%%%%%%%%%%%%%%%%%%%%%%%%%%%%%%%%%%%%%%%%%
\subsubsection{Shock Heating of Planetary Wakes}
Waves induced by a planet can also warm up the inner disk region as well as the gravitational instability \citep{Lyra2016On-Shocks-Drive,Rafikov2016Protoplanetary-,Ziampras2020The-impact-of-p}.
\citet{Ziampras2020The-impact-of-p} showed that in an optically thick disk, the shock heating of wakes generated by a (sub-)Jupiter mass planet warms up the inner region and thereby shifts the snow line outward.
For instance, when $ \dot{M} = 10^{-8} \Mpyr $ and $\alpha = 10^{-3}$, a Jupiter-mass planet at $r = 4$ au provides enough heat to raise the temperature at 1 au to 300 K (see Figure 6 in their paper).

The important point is that the perturber must have sufficient mass.
According to \citet{Ziampras2020The-impact-of-p}, the perturber with $\gtrsim$ 100 M$_{\earth}$ provides heat that increases the temperature to $\gtrsim$ 200 K at $\sim$ 1 au. 
Therefore, the proto-Jupiter needs to have formed early in order for this heating to affect our conclusion that the snow line would have passed 1 au within 1 Myr.
Interestingly, if Jupiter formed early, the oversupply of water to the terrestrial planets may also be resolved (see Section \ref{ssec:scenarios}).

\subsection{Effects of the Dust Model on Accretion Heating}

In this paper, we have assumed that the dust model is as 0.1 $\mu$m-sized grains and that the opacity is constant. 
As seen in Sections \ref{sssec:SL-opacity} and \ref{sssec:SL-fdg}, the efficiency of accretion heating in the MHD model depends on the ionization fraction and opacity.
We discuss how the dust model can be changed, and how the change will affect the efficiency of accretion heating.
%location of the snow line.  

\subsubsection{Dust Evolution}

Coagulation of dust grains alters the dust spatial and size distributions.
Especially, the abundance of smaller grains in the disk upper layer mainly determines the opacity and ionization fraction, and thus $\tau_{\rm heat}$.
The dust growth reduces the abundance of smaller grains \citep[e.g.,][]{Birnstiel2010Gas--and-dust-e,Birnstiel2011Dust-size-distr}. 
In addition, 
grains settle toward the midplane, reducing the dust abundance in the upper layer \citep{Weidenschilling1980Dust-to-planete}. 
%Besides, the larger grains also drift radially inward, which may decrease the dust surface density \citep[e.g.,][]{Birnstiel2012A-simple-model-}. 

The opacity in the upper layer is decreased by these effects of the dust evolution.
On the other hand, the effects increase the ionization fraction because the total grain surface area is decreased.
The competition between increasing opacity and decreasing ionization fraction
will determine whether the dust evolution promotes or further suppresses the disk heating. 
This will be investigated in our future work.

\subsubsection{Sublimation of Ice}

We have assumed that the opacity is constant value, but the opacity is decreased at the snow line by sublimation of ice, which affects the radial temperature profile.  
\citet{Oka2011Evolution-of-Sn} investigated the effect for the conventional viscously heated disk model
by adopting a dust opacity model of \citet{Miyake1993Effects-of-part}. 
They showed that the opacity is decreased by a factor of 3 by sublimation of ice, 
and consequently the temperature inside the snow line is decreased. 

On the other hand, the total surface area of grains is decreased by the sublimation, resulting in increasing the temperature in the MHD model (Section \ref{sssec:SL-fdg}). 
Using the mass fraction of ice and silicate in \citet{Miyake1993Effects-of-part}, 
the total surface area is decreased by a factor of 9.
Considering both effects, 
if $\tau_{\rm heat} \propto f_{\rm dg}$, 
as it is when $f_{\rm dg}$ is changed from 0.01 to 0.001 (Section \ref{sssec:SL-fdg}),
$\tau_{\rm heat}$ is increased by a factor of $\approx 3$ by the ice sublimation.
Thus, in the MHD model, the temperature may increase inside the snow line, which is different from the 
conventional models. 
In this case, the snow line can be shifted outward by 1.3 times when accretion heating determines the snow line location.
Nevertheless, this effect is negligible when irradiation heating is dominant as in the fiducial case.

%%%%%%%%%%%%%%%%%%%%%%%%%%%%%%%%%%%%%%%%%%%%%%%
%%%%%%%%%%%%%%%%%%%%%%%%%%%%%%%%%%%%%%%%%%%%%%%
%%%%%%%%%%%%%%%%%%%%%%%%%%%%%%%%%%%%%%%%%%%%%%%
\section{Summary and Conclusions}\label{sec:Conclusions}

We have investigated the migration of the water snow line in PPDs whose accretion is controlled by laminar magnetic fields, which have been proposed by various nonideal MHD simulations. 
\citetalias{Mori2019Temperature-Str} showed that the accretion heating is much less efficient than in the conventional viscous disk model.
This is because the heating in the MHD model occurs at a high altitude and also because a substantial fraction of accretion energy released is removed by the disk wind.
We proposed the empirical model of the disk temperature based on \citetalias{Mori2019Temperature-Str} to calculate the snow line location over the disk evolution (see Section \ref{sec:methods}). 

The snow line in the MHD disk model reaches the current Earth's orbit earlier than in the viscous disk model.
For instance, in our fiducial model, the time where the snow line in the MHD model passes 1 au is \tcrossMHD Myr, whereas that for the viscous model is \tcrossvisc Myr (see Section \ref{ssec:SL-fid}).
In the MHD model, the migration of the snow line is mainly driven by the temporal evolution of the stellar luminosity in the pre-main sequence phase. 
Our parameter study shows that the important parameters for efficient accretion heating are the disk ionization level and opacity (see Section \ref{ssec:SL-pdep}).
High opacity and ionization fraction lead to efficient accretion heating.
%% Add
This indicates that treating the opacity and ionization fraction consistently using an identical dust model is crucial for properly evaluating the efficiency of Joule heating.
When the opacity is scaled with the dust-to-gas ratio, effects of these parameters on the temperature will be cancelled out, and thus the accretion heating is inefficient, which is consistent with \citetalias{Mori2019Temperature-Str}.
Besides, considering that the present results are based on the maximum estimate of the disk temperature, we would expect that the accretion heating is not dominant.
Thus, the snow line in the magnetically accreting laminar disks would pass inside the current Earth's orbit within 1 Myr after star formation,  
while the time in the viscous disk model is much longer than 1 Myr.

The early arrival of the snow line to the current rocky planet orbits constrains the formation process of the planets (see Section \ref{ssec:scenarios}).
The terrestrial planets in the solar system should have formed inside the snow line because the water contents are significantly lower than that of icy bodies in the outer solar system. 
If we assume that the terrestrial planets formed at the current orbits, the protoplanets should have formed in the early phase of the disk evolution ($t < 1$ Myr).
On the other hand, if we consider the possibility that the protoplanets moved after formation, they could have formed near the star and then moved outward to the current orbits.

We should note that the present results are based on the assumption that the magnetic field drives both the disk accretion and heating.
If hydrodynamic instabilities generate the turbulence around the midplane, its energy dissipation can further warm the disk up.
Further research is needed to assess the validity of this possibility.

\acknowledgments 
The authors thank the anonymous referee for their helpful comments.
The authors also thank Shigenobu Hirose, Ogihara Masahiro, Takahiro Ueda, Yuhito Shibaike, Shigeru Ida and Ryota Fukai for fruitful discussion and meaningful comments.
This work was supported by JSPS KAKENHI Grant Numbers JP16K17661, JP17J10129, JP18H05222, JP18H05438, JP19K03926, JP19K03941, JP20H00182, JP20H00205, JP20H01948, and 21J00086, the National Key R\&D Program of China (No. 2019YFA0405100), and the joint Tohoku-Tsinghua Collaborative Research Grant.
%Numerical computations were carried out on Cray XC50 at Center for Computational Astrophysics, National Astronomical Observatory of Japan.

\appendix

%%%%%%%%%%%%%%%%%%%%%%%%%%%%%%%%%%%%%%%%%%%%%%%
%%%%%%%%%%%%%%%%%%%%%%%%%%%%%%%%%%%%%%%%%%%%%%%
%%%%%%%%%%%%%%%%%%%%%%%%%%%%%%%%%%%%%%%%%%%%%%%
\section{Vertical Temperature Structure in Internally Heated Disks} \label{app:Tderiv}
Here we derive the vertical temperature profile of an internally heated gas disk.
For simplicity, we assume disk structure that is symmetric with respect to $z = 0$,
and thus half of the heating energy is radiated from one side of the disk surfaces.
Under this assumption, the vertical temperature profile $T_{\rm acc,\,MHD}(z)$ for given heating rate per unit volume $q(z)$ can be written as \citep[][\citetalias{Mori2019Temperature-Str}]{Hubeny1990Vertical-struct},
\begin{equation}\label{eq:T}
	T_{\rm acc} (z) = \pr{ \frac{3\Gamma }{8\sigma} }^{1/4} \left[   \tau_{\rm eff}(z)
					+    \frac{  1 }{\sqrt{3}}   + \frac{2 q(z) }{3  \rho(z) \kappa_{\rm P}(z) \Gamma }  \right] ^{1/4} ,
\end{equation}
where
\begin{equation}
	\Gamma = \int_{-\infty}^{+\infty} q(z) \dz ,
\end{equation}
is the heating rate per unit area,
\begin{equation} \label{eq:taueff}
	\tau_{\rm eff}(z) = \frac{2}{\Gamma} \int_{z}^{+\infty} \rho(z')\kappa_{\rm R}(z')  \mathcal{F}(z') \dz',
	\label{eq:tau_heat}
\end{equation}
is the optical depth from $z' = z$ to $z' = \infty$ 
weighted by the radiative energy flux ${\cal F}$,
and $\kappa_{\rm R}$ ad $\kappa_{\rm P}$ are the Rosseland and Planck mean opacities, respectively.
The radiative energy flux is related to $q$ as $\partial{\cal F}/\partial z = q$; assuming that $q(z)$ is symmetric about the disk midplane, we have
\begin{equation}
	\mathcal{F}(z) = \int_{0}^{z} q(z') \dz'.
\end{equation}
The third term in the brackets of \eqref{eq:T} is negligible except at locations where $\tau_{\rm eff}(z)$ is small and $q(z)$ is anomalously high (e.g., at current sheets well above and below the midplane).  
If the heating rate scales with density as in the standard viscous model, $q(z) = (\Gamma/\Sigma)\rho(z)$, and hence the third term becomes $2/(3 \kappa_{\rm P} \Sigma)$.
When the Planck-mean optical depth at the midplane $\kappa_{\rm P} \Sigma$ is $\gg 1$, this term is much smaller than the second term for all $z$. 

Neglecting the third term in the right-hand side of \eqref{eq:T}, the temperature at the midplane can be  approximately written as
\begin{equation}\label{eq:Tmid}
	T_{\rm acc} (z=0) = \pr{ \frac{3\Gamma }{8\sigma} }^{1/4} \left(   \tau_{\rm heat}
					+    \frac{  1 }{\sqrt{3}}  \right) ^{1/4} ,
\end{equation}
where we have defined 
\begin{equation}
	\tau_{\rm heat} \equiv \tau_{\rm eff}(z=0) .
\end{equation} 
If we further assume that $\kappa_{\rm R}$ is constant,  $\tau_{\rm heat}$ can be rewritten as $\tau_{\rm heat} = \kappa_{\rm R} \Sigma_{\rm heat}$, where
\begin{equation} \label{eq:Sigmaheat}
	\Sigmaheat \equiv \frac{2}{\Gamma} \int_{0}^{\infty} \rho(z')  \mathcal{F}(z') \dz' .
\end{equation}

When accretion heating occurs only above an altitude $z_{\rm heat}$, $\Sigma_{\rm heat}$ approximately represents the mass column depth to $z  = z_{\rm heat}$ because ${\cal F}(z) = 0$ at $|z| < z_{\rm heat}$.  
%When accretion heating occurs only at $|z| > z_{\rm heat}$, $\Sigma_{\rm heat}$ approximately represents the mass column depth to $z  = z_{\rm heat}$ because ${\cal F}(z) = 0$ at $|z| < z_{\rm heat}$.  
In the extreme case where the region of accretion heating is infinitesimally thin, $\Sigma_{\rm heat}$ is exactly equal to the depth to the layer  (see Appendix A of \citetalias{Mori2019Temperature-Str}). 
For these cases, one can show that  $T_{\rm acc}(z)$ is constant at $
|z| < z_{\rm heat}$.

When accretion heating is vertically uniform in the sense that $q(z) = (\Gamma/\Sigma) \rho(z)$, we have ${\cal F} = (\Gamma/\Sigma) \chi (z)$ with $\chi(z) \equiv \int_0^z \rho(z')dz'$, which gives $\Sigma_{\rm heat} = (2/\Sigma) \int_0^\infty \chi(z) \rho(z) dz
= (2/\Sigma) \int_0^{\Sigma/2} \chi d\chi = \Sigma/4$ and 
\begin{equation} \label{eq:tauheat-visc}
	\tau_{\rm heat} = \tau_{\rm mid}/2 .
\end{equation}

\section{Energy Balance in Wind-Driven Accretion Disks}\label{app:en-bal}
The energy balance in wind-driven accretion disks is expressed as \citep[][see the equation above their Equation~(B.10)]{Suzuki2016Evolution-of-pr}
\begin{eqnarray}\label{eq:eneg-eq}
	L_{z} + \Gamma 
	&=& r\Omega w_{\rm wind} ,
\end{eqnarray}
where 
$L_{z}$ is the energy flux carried away by the wind material and $w_{\rm wind}$ is the $z \phi$ component of the total Maxwell stress exerted by the wind on both sides of the disk.
Here we assume that the disk accretion is driven only by the angular momentum transport to the disk wind, and neglect the energy liberated by the radial stress.
The right-hand side of the Equation shows the total liberated energy in the wind-driven accretion disks, while the left-hand side shows how the liberated energy is used.
The wind stress $w_{\rm wind}$ is related to the wind-driven accretion rate $\dot{M}$ as 
\begin{equation}
    \dot{M} = \frac{4\pi r}{\Omega} w_{\rm wind} .
\end{equation}
Eliminating $w_{\rm wind}$, we obtain
\begin{equation} \label{eq:wind-lib-ene}
	L_{z} + \Gamma = \frac{\Omega^{2}\dot{M}}{4 \pi} .
 \label{eq:eneg2}
\end{equation}
Now we define 
\begin{equation}
    f_{\rm heat} = \frac{\Gamma}{L_z+\Gamma}
\end{equation} 
as the fraction of heat deposited inside the disk in the total liberated energy.
In wind-driven accretion disks, using \eqref{eq:wind-lib-ene}, we have 
\begin{equation}
    \Gamma = f_{\rm heat} \frac{\Omega^2 \dot{M}}{4\pi} .
\end{equation}

The fraction $f_{\rm heat}$ depends on the polarity of the net vertical magnetic field. 
The ranges of $f_{\rm heat} $ in the MHD simulations of \citetalias{Mori2019Temperature-Str} are 0.02--0.5 and 5$\times 10^{-4}$--0.1 
when the vertical magnetic fields are aligned and anti-aligned with the disk rotation axis, respectively.
Note that the total liberated energy includes the contribution from the radial stress, though it is not dominant. 
How much the energy is dissipated correlates with the toroidal field strength.  
Hall effect amplifies the toroidal field for the aligned field case, while Hall effect damps the toroidal field for the anti-aligned case.

%%%%%%%%%%%%%%%%%%%%%%%%%%%%%%%%%%%%%%%%%%%%%%%
%%%%%%%%%%%%%%%%%%%%%%%%%%%%%%%%%%%%%%%%%%%%%%%
%%%%%%%%%%%%%%%%%%%%%%%%%%%%%%%%%%%%%%%%%%%%%%%
\section{Depth of the Heating Layer} \label{app:z_heat}

\begin{figure}[t]
	\centering
	\includegraphics[width=\hsize,clip]{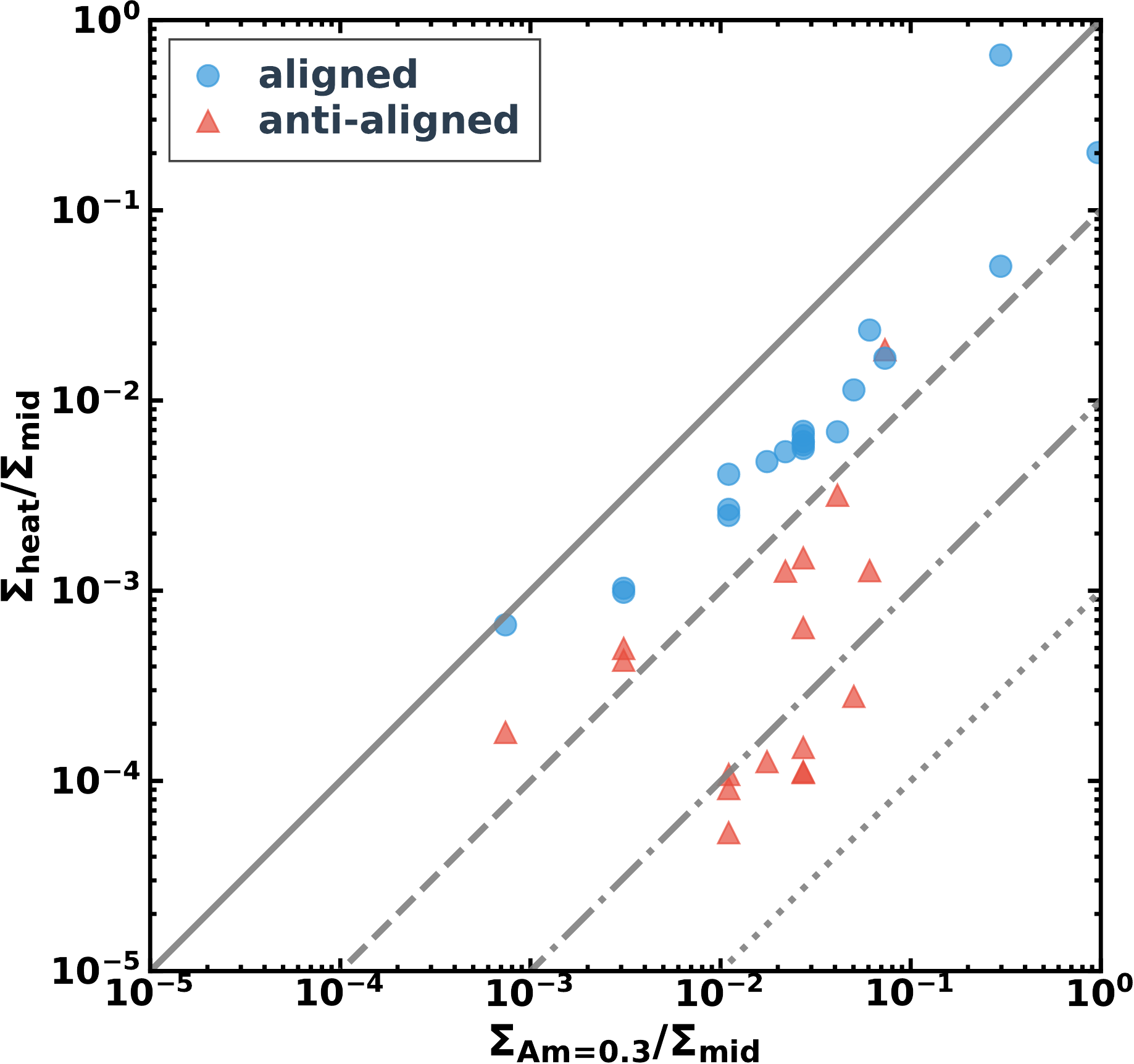}	
	\caption{
	The flux-weighted column density $\Sigma_{\rm heat}$ (see \eqref{eq:Sigmaheat}), 
	against the column depth $\SigmaAm$ at the altitude where ${\rm Am} = 0.3$, both of which are
	normalized by the column depth at the midplane, $\Sigmamid = \Sigma/2$.
	The data are given from the MHD simulation \citepalias{Mori2019Temperature-Str} and additional simulations for $f_{\rm dg}=0.01$ and various X-ray models.
	Circles and triangles correspond to the cases where the background magnetic field is 
	aligned and anti-aligned with the disk rotation vector, respectively.
	The solid, dashed, dash-dotted, and dotted lines show $f_{\rm depth} =$ 1, 0.1, 0.01, and 0.001, respectively.
		}
	\label{fig:mod-vs-res}
\end{figure}

We here show that $\SigmaAm$ gives the maximum $\Sigmaheat$. 
We use the results of nonideal MHD simulations by 
\citetalias{Mori2019Temperature-Str} that
take into account nonideal MHD effects (Ohmic diffusion, Hall effect, and ambipolar diffusion). The resistivities corresponding to the three nonideal effects are computed from the ionization balance in the disk (see Section 2.1 in \citetalias{Mori2019Temperature-Str} and Section \ref{ssec:heat-depth} for the detail of the ionization model).
We also use data from new simulations with $f_{\rm dg} = 0.01$ and with different X-ray models, 
where we vary the X-ray temperature between $\{3, 5\}$ keV and the X-ray luminosity between $ \{0.3, 1, 3\}\times 10^{30}$ erg s$^{-1}$.
All simulations are carried out for two cases where the vertical magnetic field $B_{z}$ threading the disk (i.e., background magnetic field) is set to parallel/anti-parallel to the disk rotation axis ($B_{z} >0$ and $B_{z} < 0$, respectively), because Hall effect affects the magnetic field profile depending on the polarity of the background field.

Figure \ref{fig:mod-vs-res} shows $\Sigmaheat$ versus $\SigmaAm$  for all simulation data.
Both axes are normalized by the midplane column depth $\Sigmamid = \Sigma/2$.
Different data points correspond to simulations with different parameter sets.
We also define $f_{\rm depth} = \Sigma_{\rm heat}/\SigmaAm$ (\eqref{eq:f-depth}) and plot lines of $f_{\rm depth}=1, 0.1, 0.01$ and $0.001$.
We find that all except one data point are below the line $f_{\rm depth}=1$.
The outlier corresponds to the simulation for $r = 5 $ au and $B_{z} >0$.
In this particular simulation, the midplane is ionized enough to sustain a strong current sheet \citep{Bai2015Hall-Effect-Con} that gives substantial heating. 
In this paper, we mainly focus on the region where $r \la 5 ~\rm au$, where ${\rm Am} \lesssim 0.1$ at the midplane.
For this case, we can safely use $\SigmaAm$ as the upper limit of $\Sigmaheat$.

Owing to the property of the Hall effect, the actual value of $f_{\rm depth}$ strongly depends on the direction of the background magnetic field relative to the disk rotation axis.
When $B_{z} >0$, the range of $f_{\rm depth} $ is 1--0.1, which means that the current layers lie approximately just above the ambipolar dead zone. 
This is consistent with the expectation that the ambipolar diffusion determines the altitude of the current layer (see Section \ref{ssec:heat-depth}).
When $B_{z} <0$, $f_{\rm depth} = $ 0.1--0.01. 
This is because the suppression of the magnetic field by the Hall effect acts to push the heating layer to higher altitude.
Thus, $f_{\rm depth} = 1$ provides the upper limit of the midplane temperature determined by MHD accretion heating.

\bibliography{bibfile}
\bibliographystyle{aasjournal_1}

\end{document}